\documentclass[twocolumn,prd,nofootinbib,preprintnumbers]{revtex4}

\pdfoutput=1

\usepackage{amsmath,amssymb}
\usepackage{graphicx, graphics, color}
\usepackage{slashed}
\usepackage{hyperref}
\usepackage{epsf,verbatim}
\usepackage{multirow}
\usepackage{array}
\usepackage{upgreek}
\newcommand{\fref}[1]{Fig.~\ref{f.#1}}
\newcommand{\eref}[1]{Eq.~(\ref{e.#1})}

\newcommand{\sref}[1]{Section~\ref{s.#1}}

\newcommand{\tref}[1]{Table \ref{t.#1}}

\newcommand{\beq}{\begin{eqnarray}}
\newcommand{\eeq}{\end{eqnarray}}
\newcommand{\beqs}{\begin{eqnarray*}}
\newcommand{\eeqs}{\end{eqnarray*}}

\newcommand{\ca}{{\mathcal A}}

\newcommand{\cj}{{\mathcal J}}
\newcommand{\cl}{{\mathcal L}}
\newcommand{\cm}{{\mathcal M}}
\newcommand{\co}{{\mathcal O}}

\newcommand{\cu}{{\mathcal U}}

\newcommand{\der}[1]{\mathrm{d}#1\,}

\newcommand{\muh}{\mu_\text{h}}
\newcommand{\muf}{\mu_\text{f}}
\newcommand{\mus}{\mu_\text{s}}
\newcommand{\mur}{\mu_\text{r}}
\newcommand{\Gcusp}{{\Gamma^{\text{cusp}}_\mathrm{F}}}
\newcommand{\GF}{\Gamma^{\mathrm{F}}}
\newcommand{\GA}{\Gamma^{\mathrm{A}}}
\newcommand{\gF}{\gamma^{\mathrm{F}}}
\newcommand{\ptv}{p_\mathrm{T}^\text{veto}}

\begin{document}

\title{A New Perspective on Scale Uncertainties for Diboson Processes}

\author{Prerit Jaiswal}

\affiliation{\it Department of Physics, Florida State University, Tallahassee, FL 32306, USA \\ 
                and \\
                Department of Physics, Syracuse University, Syracuse, NY 13244, USA}

\begin{abstract}
The electroweak diboson production cross-sections are known to receive large radiative corrections beyond 
leading-order (LO), approaching up to $\sim 60\%$ at next-to-leading order (NLO), compared to the scale
uncertainties which are in the range $1$-$5\%$ at LO. If the scale uncertainties are to be taken seriously, 
the NLO predictions are as much as $\sim 30 \sigma$ away from their LO counterpart suggesting a very poor
convergence of the perturbation theory. In this paper, we show that there is a 
second source of scale uncertainty which has not been considered in the literature, namely 
the complex phase of the scales, which can lead to large perturbative corrections. Using the 
formalism of soft-collinear effective theory,  we resum these large contributions from the complex phase, 
finding that the scale uncertainties in fixed-order calculations can be grossly underestimated compared to the 
resummed predictions, which have uncertainties as large as $13$--$16\%$ at LO. Even at NLO,
we find that the scale uncertainties are marginally higher than previously estimated, depending on the choice of 
scale. Using our method of scale variation, the compatibility of LO and NLO results within the scale uncertainties
 is vastly improved so that the perturbation theory can be relied upon. This method of scale variation can be easily 
 extended to beyond NLO calculations as well as other LHC processes. 

%revisit pair-production of heavy electroweak gauge bosons and show that nearly a half of these NLO corrections can be 
%accounted by $\co(\alpha_s \pi^2) $ terms, which appear in higher order calculations. Using the formalism of 
%soft-collinear effective theory, we resum such $\pi^2$ terms to all orders and find an improved convergence in 
%perturbation theory in going from LO to NLO. The impact of $\pi^2$ resummation beyond NLO is an increase 
%by $7\%$, $9\%$ and $11\%$ for $ZZ$, $W^+W^-$ and $W^\pm Z$ channels respectively at $\sqrt{s}=7$--$14$ 
%TeV LHC runs.  We also resum the $\pi^2$ terms for purely gluon-induced $W^+W^-$ and $ZZ$ processes 
%which leads to a further $2$--$4\%$ increase over the combined NLO$+gg$ total cross-section. 
\end{abstract}

\maketitle

%%%%%%%%%%%%%%%%%%%%%%%%%%%%%%%%%%%%%%%%%%%%%%%%%%%%%%%%%%%%%%%%%%%%%%%%%%%%
%%%%%%%%%%%%%%%%%%%%%%%%%%%%%%%%%%%%%%%%%%%%%%%%%%%%%%%%%%%%%%%%%%%%%%%%%%%%

\section{Introduction}
A precise understanding of the electroweak gauge boson pair-production at the LHC is critical for several reasons. 
First and foremost, many of the diboson processes are dominant backgrounds to Higgs production and its 
subsequent decays to Standard Model (SM) particles. A good understanding of the diboson background is 
therefore crucial in the measurement of the Higgs couplings to the SM particles. Secondly, diboson processes
constitute an important test for the electroweak sector. And finally, diboson processes are often backgrounds to many new physics 
processes, making it challenging to distinguish one from the other.  

In this paper, we focus on heavy electroweak vector boson pair-production channels, $W^+W^-$, $ZZ$ and 
$W^\pm Z$, owing to their similar kinematics. The cross-sections measured by the ATLAS 
\cite{ATLAS:WW8, ATLAS:ZZ8, ATLAS:WZ8, ATLAS:WW7, ATLAS:ZZ7, ATLAS:WZ7} and the 
CMS \cite{CMS:WW8, CMS:ZZ8, CMS:WZ7and8, CMS:WW7, CMS:ZZ7} collaborations in these channels at 
$\sqrt{s} = 7$ TeV  and $8$ TeV  LHC runs are compatible with the theory predictions within $2\sigma$. Three 
measurements where the discrepancy exceeds $1\sigma$ level are the $W^\pm Z$ measurements by  CMS 
 and the $W^+ W^-$ measurements by both ATLAS and CMS collaborations. The discrepancy in 
 the $WW$ channel is particularly compelling given that both ATLAS and CMS experiments observe an excess of
 $\sim 20\%$ over the SM theory prediction, which has fueled speculations that new physics could be 
 hiding in the $W^+W^-$ measurements \cite{NewPhyWW:1, NewPhyWW:2, NewPhyWW:3, NewPhyWW:4, 
 NewPhyWW:5, NewPhyWW:6, NewPhyWW:7, NewPhyWW:8}. In order to test the possibility of new physics  
 mimicking the SM background, a precise theoretical understanding of the higher-order corrections to the SM 
 diboson production is essential.  
 
 The study of higher-order corrections to diboson production has a long history, with the first NLO QCD 
 corrections to $W^+W^-$, $ZZ$ and $W^\pm Z$ channels computed in  \cite{WW:NLO1, WW:NLO2}, 
 \cite{ZZ:NLO1, ZZ:NLO2} and \cite{WZ:NLO}, respectively. Leptonic decays of dibosons without 
 spin-correlations was studied in \cite{VV:Leptonic}. One-loop helicity amplitudes for leptonic decays of 
 vector boson pair were computed in \cite{VV:Helicity}, allowing for complete NLO computation in
 \cite{VV:NLOfull1, VV:NLOfull2}. The $W^+ W^-$ and $ZZ$ cross-sections also receive contributions from 
 the gluon-fusion channel, which although formally NNLO, can be significant owing to large gluon parton 
 distribution functions (PDFs) at the LHC. These corrections were calculated in \cite{ggVV:1, ggVV:2} with 
 the corresponding leptonic decays included in \cite{ggZZ:lep1, ggZZ:lep2, ggZZ:lep3, ggWW:lep1, ggWW:lep2}.
 The complete NLO calculations including leptonic decays, spin-correlations and gluon-fusion contributions,
 for all diboson channels, was presented in  \cite{VV:NLOfull+gg}. Recently, electroweak calculations have also 
 been considered For $W$ pair-production \cite{WW:EW1, WW:EW2}, and for $ZZ$ and $W^\pm Z$ production 
\cite{VV:EW1, VV:EW2}.  NLO QCD corrections to $W^+ W^-$  and $ZZ$ production with one jet  have been 
computed in  \cite{WW+1jet:1, WW+1jet:2, WW+1jet:3} and \cite{ZZ+1jet}, respectively, while $W^+ W^- + 2$
jets calculations were considered in \cite{WW+2jets:1,WW+2jets:2}. Transverse momentum resummation effects
in diboson production have been studied in  \cite{VV:pT, WW:pT1, WW:pT2}, while a jet-veto study for $W^+ W^-$
channel was presented in \cite{WW:JetVeto}. The threshold corrections arising from soft-gluon resummation were 
calculated in \cite{WW:Threshold, VZ:Threshold}. Finally, the NNLO QCD corrections to $W^+ W^-$ and $ZZ$ 
have been recently computed in \cite{WW:NNLO} and  \cite{ZZ:NNLO} while the two-loop helicity amplitudes for 
all diboson channels have been calculated in \cite{VV:NNLO}.

\begin{figure*}
        \center
        \begin{tabular}{ccc}
                \includegraphics[width=0.32\textwidth]{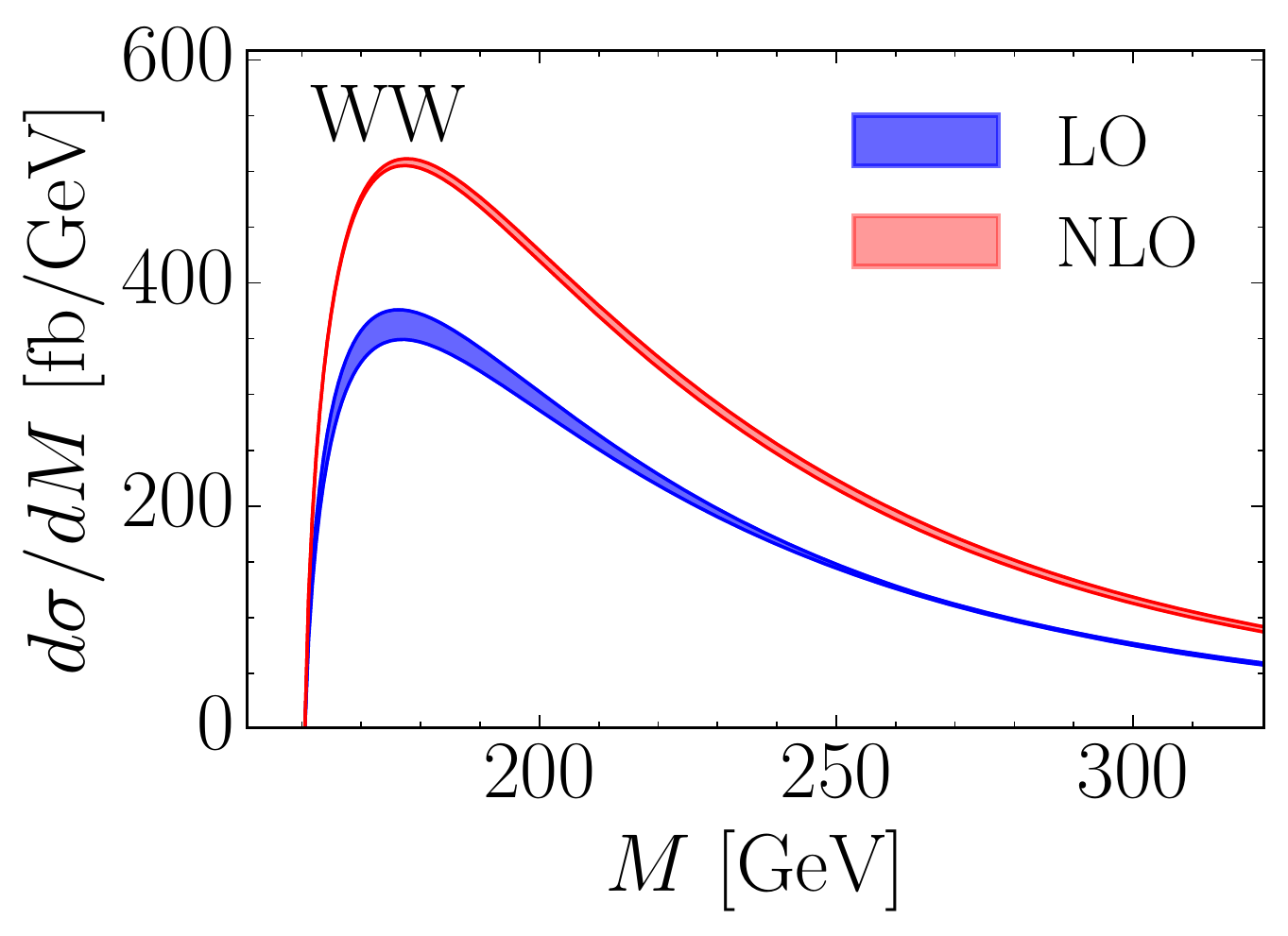}
        &
                \includegraphics[width=0.32\textwidth]{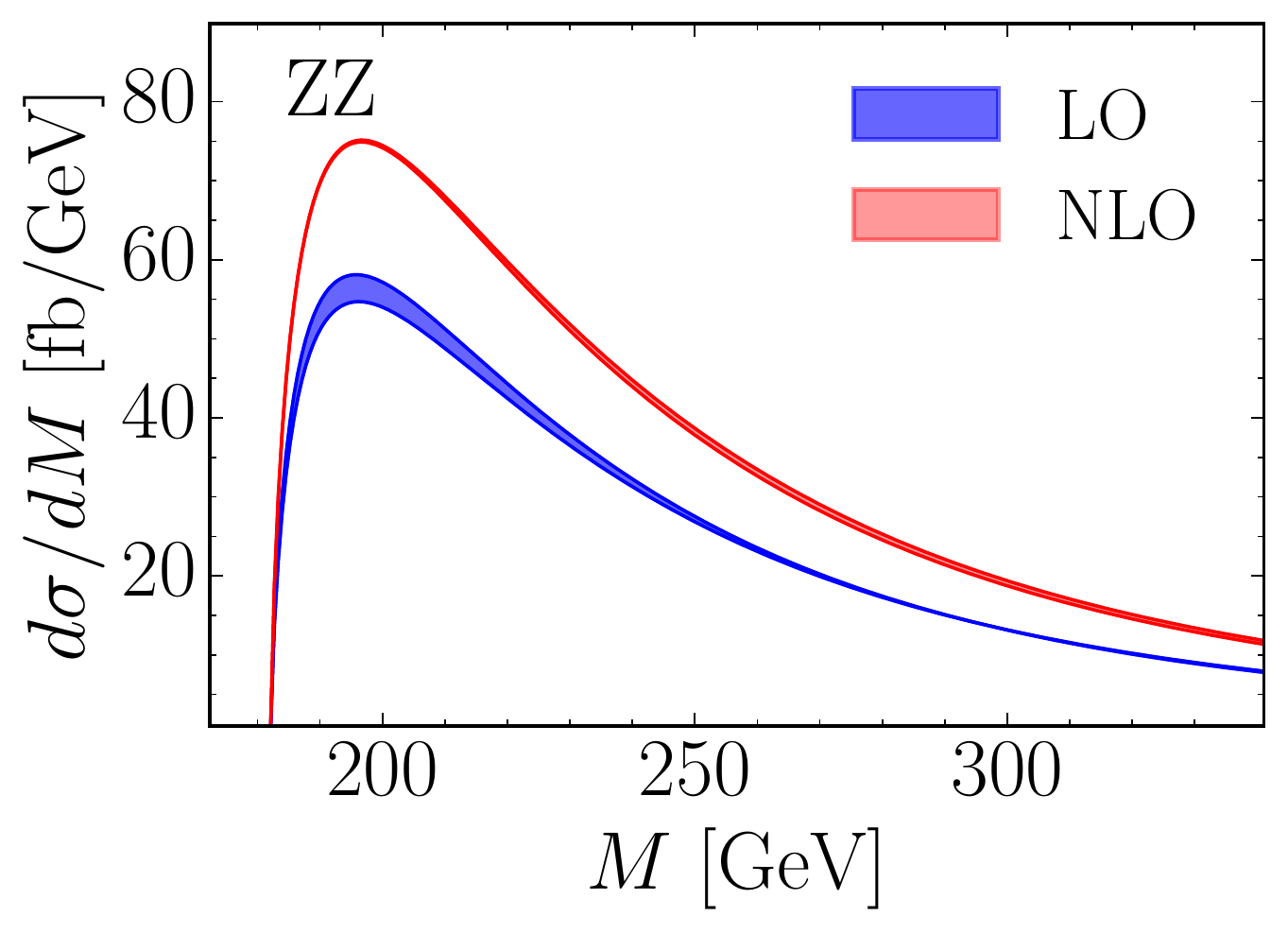}
        &
                \includegraphics[width=0.32\textwidth]{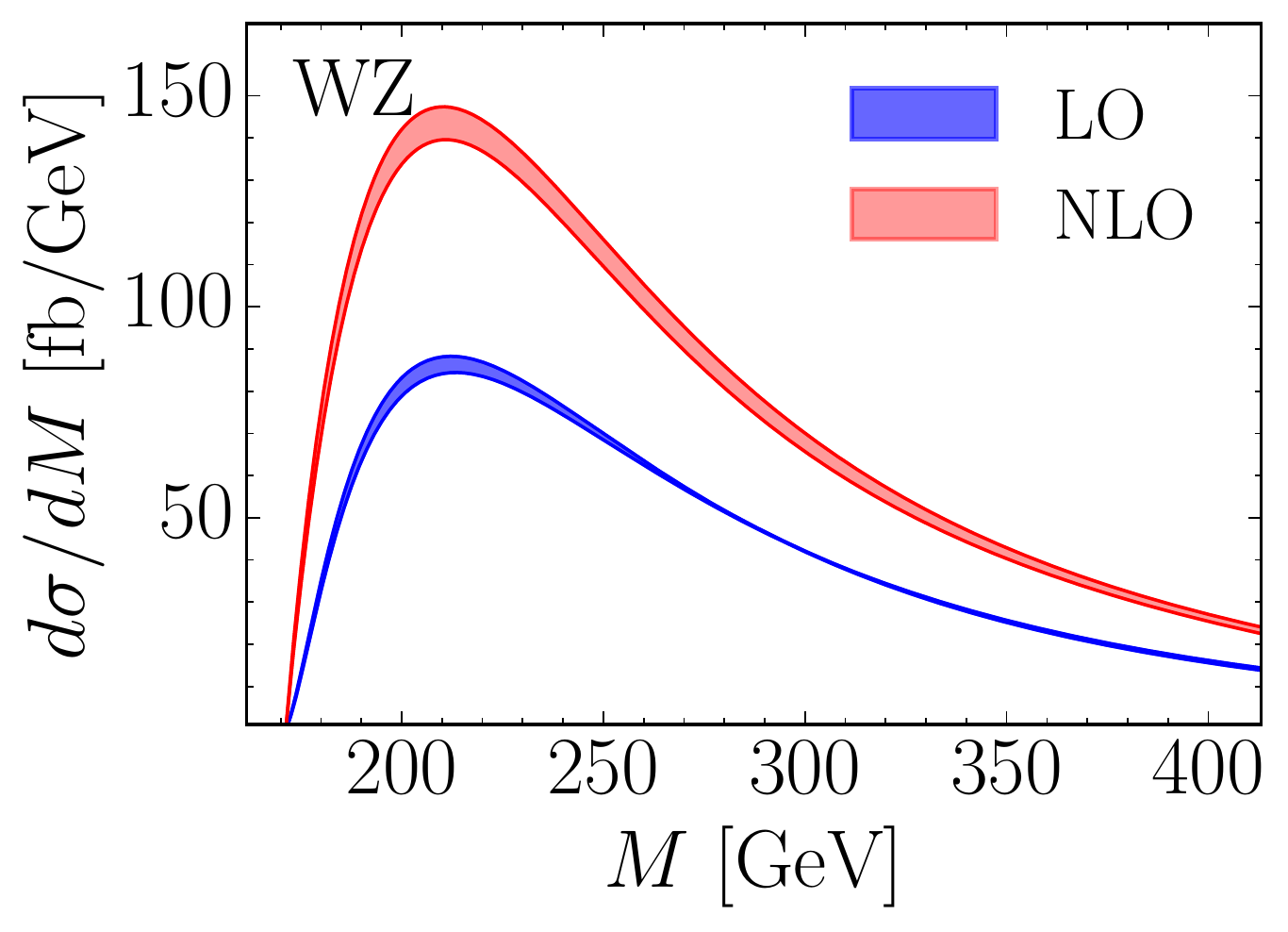}
	\\ \end{tabular}
        \caption{Differential cross-sections at LO (blue band) and NLO (red band) are shown for $W^+ W^-$ (left), 
        $ZZ$ (center) and $W^\pm Z$ (right) production at $\sqrt{s}=8$ TeV LHC run as obtained from {\tt MCFM}. 
        The uncertainty bands follow from varying the renormalization and factorization scales as $M/2 < \mu_r=\mu_f < 2 M$. 		Contributions from gluon-fusion channels are not included.}
        \label{f.8TeV_NLO}
\end{figure*}

Every higher order QCD calculation discussed above includes powers of logarithms of the form 
$\log \left[ (-M^2-i 0^+)/\mu^2 \right]$ where $M$ is the invariant mass of the diboson system and $\mu$ is the 
factorization scale, which is also the scale at which the PDFs are evaluated. Given that $\mu$ dependence of the
cross-sections is primarily controlled by the logarithmic terms, $\mu \sim M$ seems to be a reasonable choice
to minimize the higher order corrections. Further, given that physical observables are $\mu$-independent, one 
can estimate scale uncertainty in the cross-sections by varying $\mu$.  The scale uncertainties in diboson 
invariant mass distributions at LO and NLO shown in \fref{8TeV_NLO} are obtained by varying the renormalization scale, 
set equal to the factorization scale ($\mu = \mu_r = \mu_f)$, as $M/2 < \mu < 2 M$.\footnote{Varying the scale 
by factors of $1/2$ and $2$ around a central value is the standard convention followed in the literature.} 
Contrary to the naive expectations, the NLO perturbative corrections with K-factors in the range 
$1.4$--$1.7$ \cite{VV:NLOfull+gg} far exceed the scale uncertainties. If this scale uncertainty estimate is to be 
taken seriously, the NLO predictions are as much as  $\sim 30 \sigma$ away from their corresponding 
LO vales, suggesting that the perturbation theory is very poorly converging.     

We argue that there is a second source of scale uncertainty which has not been considered in the literature.
If the scale $\mu$ is allowed to be complex-valued, there is an additional parameter that must be considered
for estimating the scale uncertainties, namely the complex phase of $\mu^2$. In fact, the logarithms in the
higher-order corrections have a branch-cut along the negative real axis, so that $\mu^2 < 0$
is preferred  over $\mu^2 > 0$ to minimize logarithms. This is slightly problematic though since the PDFs 
are necessarily evaluated at $\mu^2>0$ leading to large $\pi^2$ terms when the logarithms are squared.     
 Summation of $\pi^2$ terms has been known for a long time \cite{PiSq:1, PiSq:2, PiSq:3, PiSq:4}, 
 and has been recently applied to the case of Higgs production at the LHC \cite{PiSq:Higgs}. $\pi^2$ 
 resummation calculations for diboson production have been performed in the context of threshold 
 resummation for $W^\pm Z$ and $ZZ$ channels \cite{VZ:Threshold}, and jet-veto resummation for 
 $W^+ W^-$ channel \cite{WW:JetVeto}.

The aim of this paper is to show that the variation in the phase angles of the complex renormalization scales
is essential in order to estimate the true scale uncertainties. Just as the variation in the factorization scales
is governed by the evolution of the PDFs, the variation of the phase angle will be governed by a different
renormalization group (RG) equation, which we obtain using the formalism of soft-collinear effective theory 
(SCET)  \cite{SCET:label1, SCET:label2, SCET:label3, SCET:label4, SCET:mp1, SCET:mp2} by generalizing 
the concept of {\it $\pi^2$-resummation}, where the phase angle is fixed to $(-\pi+0^+)$, to arbitrary phase angles. 
While we explicitly focus on heavy vector-boson pair production, our scale variation technique can be 
extended to any other process once its RG equation is known.\footnote{Our analysis is trivially extended 
to a class of processes which involve colorless final states such as Drell-Yan, as these processes satisfy 
the same RG equations.}   

This paper is organized as follows. In \sref{origin}, using the SCET construction for diboson production, we
demonstrate that complex-valued scales, not only arise naturally in radiative corrections, but are also associated
with large perturbative corrections. In  \sref{resum}, the large perturbative corrections arising from the complex 
phases of the scales are resummed to all orders in perturbation theory for $W^+W^-$, $ZZ$ and $W^\pm Z$ 
processes, including gluon-fusion production channels, allowing us to study scale variation for complex 
scales. Finally, in \sref{results}, our scale variation technique is applied to diboson processes, and 
numerical results for the diboson production cross-sections are presented for $\sqrt{s}=7,8,13$ and $14$ TeV 
LHC runs.

\section{Complex Scales and Large Perturbative Corrections}
\label{s.origin}

Any cross-section measurement at the LHC is characterized by a process-dependent {\it hard}-scale and one or more 
measurement-kinematics dictated {\it soft}-scale(s). For example, the {\it hard}-scale for diboson production is the invariant 
mass of the boson-pair, $M$ and the {\it soft}-scale is $\Lambda_{QCD}$ for an inclusive measurement while jet-$p_T$ 
measurements introduce another intermediate {\it soft}-scale, $p_T^{jet}$. It is well known that the presence of multiple 
scales in the theory can lead to large logarithms of the ratio of the scales, which can render the perturbation theory invalid. 
Effective field theories, on the other hand, are adept at dealing with the problem of multiple scales by renormalization 
group (RG) evolution to a single scale, effectively providing a powerful technique to resum the large logarithms. 

As already mentioned before, one such logarithm that appears in the radiative corrections to diboson 
processes is of the form 
$\alpha_s^n \log^{2n}[(-M^2 - i 0^+)/\mu^2]$, as we will explicitly show later in this section. The structure of the 
logarithm already motivates us to choose a complex-valued $\mu^2$. However, the factorization scales at which the 
PDFs are evaluated are always real valued, so that logarithms take the form $\alpha_s^n [\log(M^2/\mu^2)-i \pi]^{2n}$.  
Even for the choice of $\mu \approx M$, large perturbative corrections in the form of $\alpha_s^n \pi^{2n}$ terms remain. 
The question whether such $\pi^2$ terms should be resummed or treated as part of non-logarithmic corrections is 
a highly debated subject, which we will address later in \sref{resum}. Nonetheless, it is clear that there exists a 
hierarchy of scales in the complex $\mu^2$-plane that lead to large perturbative corrections and should be 
resummed. We will employ SCET to resum such terms in \sref{resum} but first, we set up the basic notation for diboson production in SCET 
formalism.       

Consider the inclusive vector-boson pair production, $p p  \rightarrow V V' + X$ where $V, V' \in \left\{ W, Z \right\} $ 
and $X$ is any hadronic final state. We will primarily focus on the process $q \bar{q}' \rightarrow V V'$ which is the 
only production channel at LO. Throughout this paper, we extensively follow the SCET construction and notation used
 in \cite{WW:JetVeto}, which we refer to the interested readers for more details. Let us begin by writing down the 
 SCET Lagrangian for $V V'$ production :
\beq
\cl = \frac{1}{M} \Big[ \epsilon_\mu^V \Big]^* \Big[ \epsilon_\nu^{V'} \Big]^*  
e^{i(p_V+p_{V'}) \cdot x}\cj^{\mu \nu} (x)
\label{e.SCET:Lagrangian}
\eeq  
where $\epsilon_\mu$ are the spin and momentum dependent gauge-boson polarization vectors, $p_{V}$ and $p_{V'}$ are the
gauge-boson four momenta, and $\cj$, the SCET operator 
describing the interaction of incoming `quark-jets' with the external outgoing gauge bosons, is given by
\beq
\cj^{\mu \nu} (x) = \int \der{t_1} \der{t_2} \Big[ C^{\mu \nu} (t_1,t_2,p_V, p_{V'},\mu) \nonumber \\
\times \; \chi_{\bar{c}}^\alpha (x^- + t_2) \Gamma_\alpha^{\; \beta} \chi_{c\, \beta} (x^+ + t_1) \Big] 
\label{e.SCET:current}
\eeq
Here, $\chi_c$($\chi_{\bar{c}}$) is a gauge-invariant collinear (anti-collinear) quark field in SCET
\footnote{The gauge invariance of $\chi_c$ and $\chi_{\bar{c}}$ is implemented by dressing the bare 
quark fields with collinear gluon Wilson lines \cite{SCET:label1, SCET:label2, SCET:label3}.}, 
$\Gamma$ is a spinor structure explicitly defined in \cite{WW:JetVeto} and $C^{\mu \nu}$ is the 
Wilson coefficient of the SCET operator which is a function of gauge boson momenta, the RG scale $\mu$ as well as 
the spatial parameters $t_1$ and $t_2$ along the light-cone directions as allowed by the non-locality of the SCET operators.
 In the above expression, the spinor indices have been made explicit, while the color and flavor indices are implicit. The
 multipole expansion of the operators in \eref{SCET:current}, as dictated by requiring inclusive measurements, is slightly 
 different from that in \cite{WW:JetVeto} where jet-veto condition is imposed. Nonetheless, the hard coefficients which appear 
in the factorized SCET cross-sections for $q \bar{q}' \rightarrow V V'$ production are the same in either case, and ultimately 
the only ingredients affected by the resummation in the complex $\mu^2$-plane, as we will show later. They are given by 
\beq 
C(\mu) = \tilde{C}^{\mu \nu} \left[ \tilde{C}^{\rho \sigma} \right]^* 
 \Big[ \epsilon^V_{\mu} \Big]^*  \Big[ \epsilon^{V'}_{\nu} \Big]^* \epsilon_{\rho}^V  \epsilon_{\sigma}^{V'}
\label{e.HardCoeff}
\eeq
where, $\tilde{C}^{\mu \nu}$ is the Fourier-transform of the position space Wilson coefficients $C^{\mu \nu}$ which appear
in \eref{SCET:current}. 
For brevity, here and throughout the rest of the paper, the quark flavor, helicity and momentum dependence of the 
hard coefficients will be suppressed. Also implicit are the gauge-boson spin and momentum dependence on the RHS 
of the above equation, including the summation over the final state spins. 

A typical SCET calculation for LHC observables involves computing the Wilson coefficients by matching the SCET operators
to the full QCD at a {\it hard} scale $\muh$, and then RG evolving the coefficients to a factorization scale $\muf$ at which the 
PDFs are evaluated, where the second step resums the large logarithms of the ratio $\muh/\muf$. For $VV'$ production, the 
hard coefficient at one-loop takes the following form at the matching scale $\muh$ \cite{WW:JetVeto}:  
\beq
\begin{split}
C(\muh) &= \bigg[ 1 - \frac{C_F \alpha_s (\muh)}{4 \pi} \Big( 2 L_M^2 (\muh) - 6 L_M (\muh) \\ 
& + \frac{\pi^2}{3} \Big) \bigg] \left| \cm_0 \right|^2 
+ \frac{C_F \alpha_s (\muh)}{2 \pi} \mathrm{Re} \left( \cm_0^* \cm_{1,\text{reg}} \right) 
\end{split}
\label{e.Wilson}
\eeq   
where, $L_M(\mu) = \log[ (-M^2- i 0^+)/\mu^2]$, $\cm_0$ is the Born-level amplitude for the process 
$q \bar{q}' \rightarrow V V'$, and $\cm_{1,\text{reg}}$ is the one-loop amplitude for the same process with the IR poles
subtracted using the $\mathrm{\overline{MS}}$ scheme.  As already discussed before, an optimal choice for the 
matching scale $\muh$ that minimizes the higher order corrections arising from the logarithms $L_M(\muh)$ is 
$\muh^2 \sim M^2 e^{-i (\pi-0^+)}$, rather than $\muh^2 \sim M^2$ \cite{PiSq:1, PiSq:2, PiSq:3, PiSq:4, PiSq:Higgs}. 
On the other hand, the PDFs (or more generally `beam-functions' \cite{Beam:1, Beam:2} 
for less inclusive observables) that multiply the hard coefficients in the cross-sections are typically evaluated at 
factorization scales, $\muf$ which are real valued, in contrast to our optimal choice of the matching scale 
which is phase-shifted by $\pi$ in the complex $\mu^2$-plane. Therefore, even for the case $\muf=|\muh|$, 
there exists a hierarchy of scales in the complex $\mu^2$-plane leading to large perturbative corrections that 
need to resummed, which we discuss in the next section.

\section{Resummation and Scale Variation}
\label{s.resum} 

Before we describe the resummation of large perturbative terms associated with the complex phase of $\mu^2$, 
let us define the hierarchy of scales more precisely. For a typical measurement involving $VV'$ final states, inclusive
or otherwise, there are at least two scales in the problem: the hard scale, $\muh$ and the factorization scale, 
$\muf$.\footnote{More generally, one can consider a soft scale $\mus \sim \Lambda_{QCD}$ but we assume that 
the evolution from $\mu = \mus$ to $\mu = \muf$ is accounted by the PDF running. This is true when the `threshold
corrections' from soft-emissions are small, which has been shown for the diboson processes 
\cite{WW:Threshold, VZ:Threshold}.}  Given that $\muh$ is complex-valued, the RG evolution of the
hard coefficients can be realized as a two step process, $C(\muh) \rightarrow C(|\muh|) \rightarrow C(\muf)$.
In this paper, we will consider inclusive cross-sections so that it is reasonable to set $\muf = |\muh| \equiv \mu$. 
For less inclusive measurements, such as imposing jet-veto \cite{WW:JetVeto}, we have $\muf \neq |\muh|$ so 
that the evolution $C(|\muh|) \rightarrow C(\muf)$ must also be considered. Nevertheless, the first RG running, 
$C(\muh) \rightarrow C(|\muh|)$ essentially decouples from the second RG running, $C(|\muh|) \rightarrow C(\muf)$,
so that our analysis can be trivially extended to less-inclusive measurements.   

Let us define $\upmu \equiv \muf = |\muh|$ and $\muh^2 = \upmu^2 e^{i \Theta}$, where $\Theta \in (-\pi, \pi)$ is the 
complex phase angle. In the last section, we showed that the logarithms $L_M(\muh)$ present in the hard matching 
coefficient are minimized for $\upmu=M$ and $\Theta=-\pi+0^+$. While the effective field theory dictates the choice
of the hard matching scale to be the scale of the hard interaction such that $\upmu = \co(M)$ and $\Theta = \co(-\pi)$,
there is nonetheless an ambiguity associated with the choice of the hard scale parameters, $\upmu$ and $\Theta$, 
since the contribution of non-logarithmic terms in \eref{Wilson} maybe sizable. On the other hand, total cross-section, 
being a physical observable, is independent of the choice of matching scale. Therefore, this ambiguity in the choice of 
matching scale parameters should be reflected as scale uncertainty in the theory prediction. 

Variation of the hard scale in the complex $\mu^2$-plane is shown in \fref{ComplexMu}, where the shaded annulus 
corresponds to the region $M/2 < \upmu < 2M$ and $-\pi < \Theta < \pi$.  If the non-logarithmic terms in 
\eref{Wilson} were completely dominant over the logarithmic ones, there would be no preferred value 
of $\Theta$. On the other extreme, if logarithmic terms were completely dominant, $\Theta=-\pi+0^+$
would be the ideal choice. Numerically, for the diboson processes, we find that $\pi^2$ terms arising from the 
logarithms account for nearly a half of the total NLO corrections, so that the situation is somewhere in between. 
With these considerations in mind, to estimate the scale uncertainties for diboson processes, we select the region 
$-\pi<\Theta<0$ as indicated by the green hatched region in \fref{ComplexMu}. This is to be contrasted with the 
fixed-order calculations which have $\Theta=0$ on one hand, and $\pi^2$-resummation calculations which select 
$\Theta=-\pi+0^+$ on the other hand.  

\begin{figure}
        \center
                \includegraphics[trim= 0mm 16mm 0mm 16mm, clip, width=0.4\textwidth]{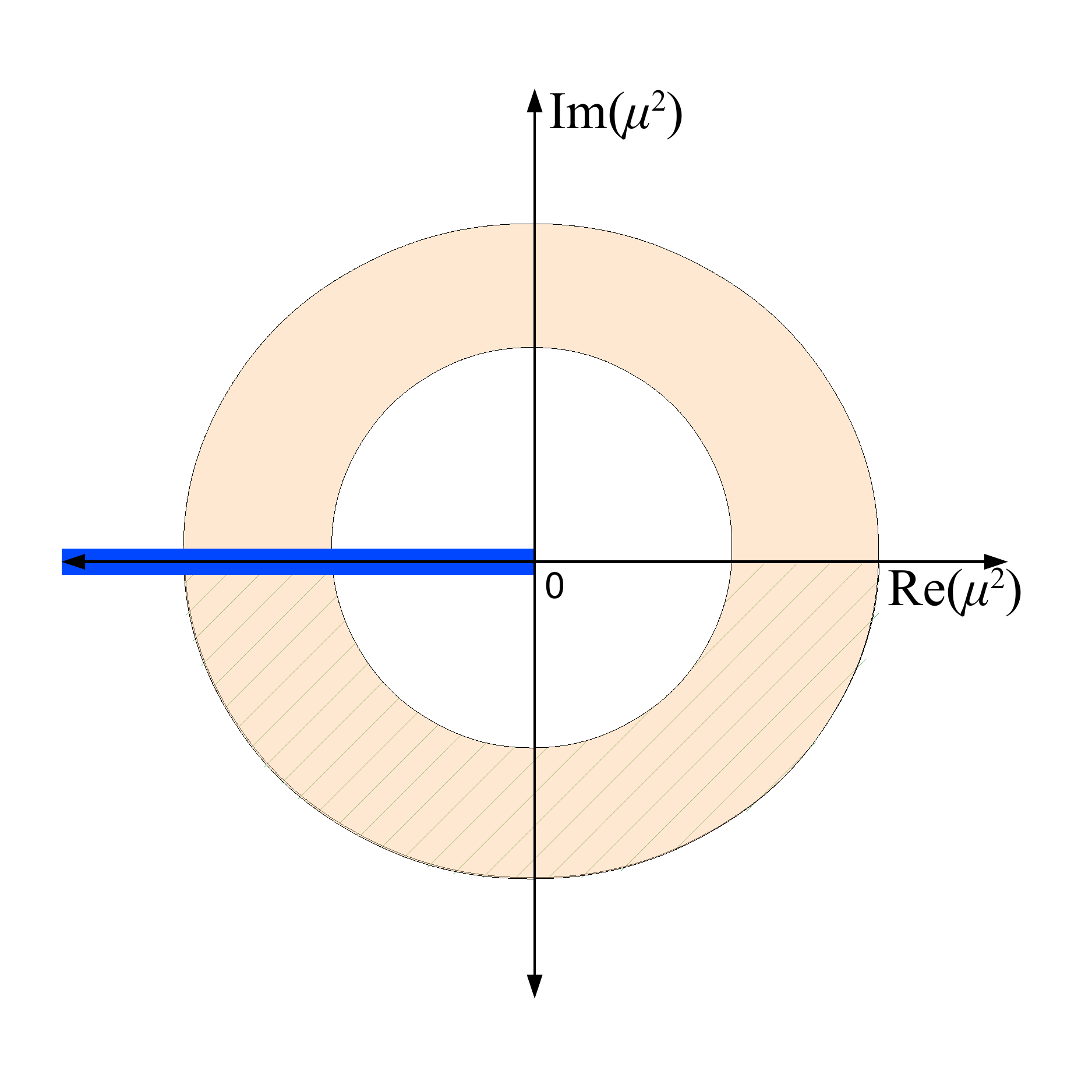}
        \caption{Variation of the hard scale $\muh$ is shown in the complex $\mu^2$-plane with a branch cut along the 
        negative real axis. The orange shaded region satisfies $M/2 < |\muh| < 2 M$ but only the hatched region of the
        annulus is considered for scale variation.} 
        \label{f.ComplexMu}
\end{figure}

For the process $q \bar{q}' \rightarrow V V'$, the scale dependence of the hard coefficients in \eref{HardCoeff} 
follows from that of the Wilson coefficients, which in turn satisfy the following RG equation :
\beq
\mu \frac{\der{\tilde{C}^{\mu \nu}(\mu)}}{\der{\mu}} = \bigg( \Gcusp L_M(\mu)  + 2 \gamma_\mathrm{F} \bigg)  
\tilde{C}^{\mu \nu} (\mu)
\label{e.RG}
\eeq
where $\Gcusp$ is the cusp-anomalous dimension which resums double logarithms while $\gamma_\mathrm{F}$ 
is the anomalous dimension which resums single logarithms. Both $\Gcusp$ and $\gamma_\mathrm{F}$ implicitly 
depend on $\mu$ through $\alpha_s$. The anomalous dimensions appearing in the RG equation above are universal 
for class of processes which have colorless final states (not counting emissions from initial state quarks),  and therefore 
identical for all diboson production processes and Drell-Yan.

A subtlety that emerges from the RG running between the scales $\muh$ and $\muf$ is that the strong coupling 
$\alpha_s(\mu)$ must now be defined in the complex $\mu^2$-plane with a branch cut along the negative real axis.  
As long as the contours of integration are sufficiently away from the Landau pole in the complex $\mu^2$-plane, 
$\alpha_s(\mu)$ is well-defined along such contours. Using the definition of QCD beta function $\beta(\alpha_s)$
and performing contour integration, a particularly useful result can be obtained \cite{PiSq:Higgs} :
\beq
\int_{\alpha_s(\upmu)}^{\alpha_s(\muh)} \frac{\der{\alpha_s}}{\beta(\alpha_s)} = \frac{i \Theta}{2} 
\label{e.Theta}
\eeq  
For the purpose of power counting in $\alpha_s$, we
shall treat $|\Theta| \sim \co(\alpha_s^{-1})$ although numerically $\Theta$ can also be zero. 
\eref{Theta} allows us to compute the complex couplings $\alpha_s(\muh)$ in terms of the real couplings  
$\alpha_s(\upmu)$, where the latter can be computed in a standard way. At NLO, we have the following relation :
\beq
\begin{split}
\frac{\alpha_s(\upmu)}{\alpha_s(\muh)} &= 1+ i a(\upmu) \frac{\Theta}{\pi} \\ 
& \quad + \frac{\alpha_s(\upmu)}{4 \pi}
\frac{\beta_1}{\beta_0} \log \left[ 1 + i a(\upmu) \frac{\Theta}{\pi} \right] + \co(\alpha_s^2) 
\end{split}
\label{e.complex_alpha}
\eeq
where $a(\upmu) = \beta_0 \alpha_s(\upmu)/4$ and $\beta_0 = 11/3 \,C_A - 4/3 \,T_F n_f$ with $C_A=4$, $T_F=1/2$ and
$n_f$ is the active number of flavors which we take to be five. Numerically, since $a(\upmu) \approx 0.2$, 
\eref{complex_alpha} is a good approximation even at NNLO. 

Having addressed the subtleties associated with complex values of $\mu^2$, we can solve the RG equation in
 \eref{RG} to evolve the Wilson coefficients from $\mu = \muh$ to $\mu = \upmu$
\beq
\tilde{C}^{\mu \nu} (\upmu) = \cu (\upmu, \muh) \tilde{C}^{\mu \nu} (\muh)
\eeq
where, the analytical expression for the evolution kernel $\cu$ can be found in \cite{WW:JetVeto, SCET:DY}. Counting 
$\Theta$ as $\co(\alpha_s^{-1})$ and neglecting $\co(\alpha_s)$ and higher-order terms in $\log \cu$, to a good 
approximation we have
\cite{PiSq:Higgs},
\beq
\begin{split}
\log \left| \cu (\upmu, \muh) \right|^2 &\approx \frac{\GF_0 \Theta^2 \alpha_s(\upmu)}{8 \pi} 
 \bigg[ 1 + \frac{\alpha_s(\upmu)}{4 \pi} \bigg\{ \\
& \quad   \frac{\GF_1}{\GF_0} -  \frac{2 \beta_0 \gF_0}{\GF_0}  
-\beta_0 \log \left( \frac{M^2}{\upmu^2} \right) \bigg\} \bigg]
\end{split}
\label{e.kernel}
\eeq  
In the above equation, $\GF_0 = 4 C_F$, $\gF_0 = - 3 C_F$, 
\beq
\GF_1 = 4 C_F \left[ C_A \left( \frac{67}{9} - \frac{\pi^2}{3} \right) - \frac{20}{9} T_F n_f \right]
\label{e.GF1}
\eeq
where, $C_F = 4/3$ and the remaining symbols have already been defined below \eref{complex_alpha}. 
It has been shown in \cite{PiSq:Higgs} that the impact of including $\co(\alpha_s)$ and higher-order terms
 in $\log |\cu|^2$ is small so that  \eref{kernel} is a good approximation even at higher-orders. Before concluding 
 this section, we briefly comment on the diboson production from the gluon-fusion channel. 

\begin{figure*}
        \center
        \begin{tabular}{ccc}
                \includegraphics[width=0.32\textwidth]{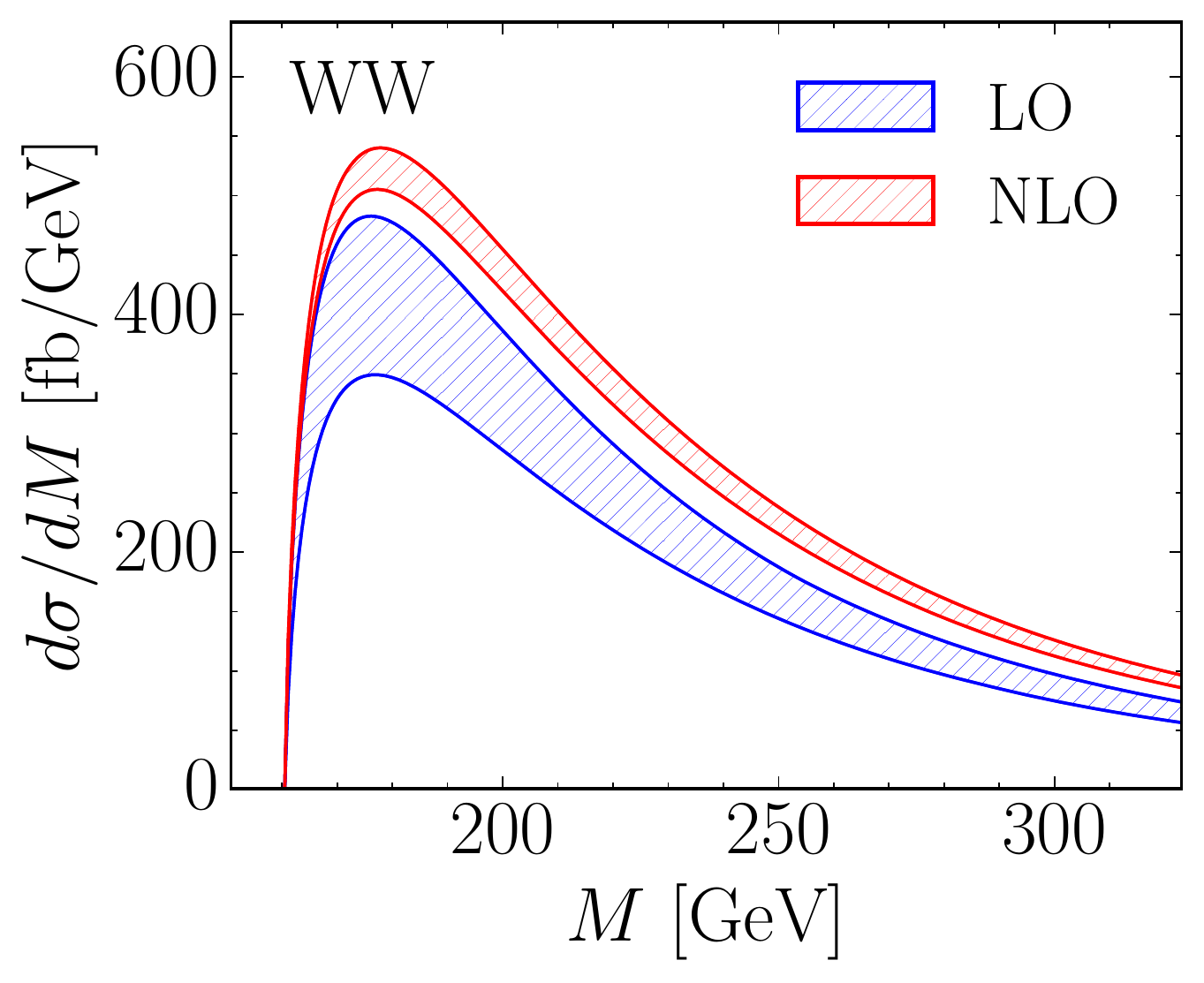}
        &
                \includegraphics[width=0.32\textwidth]{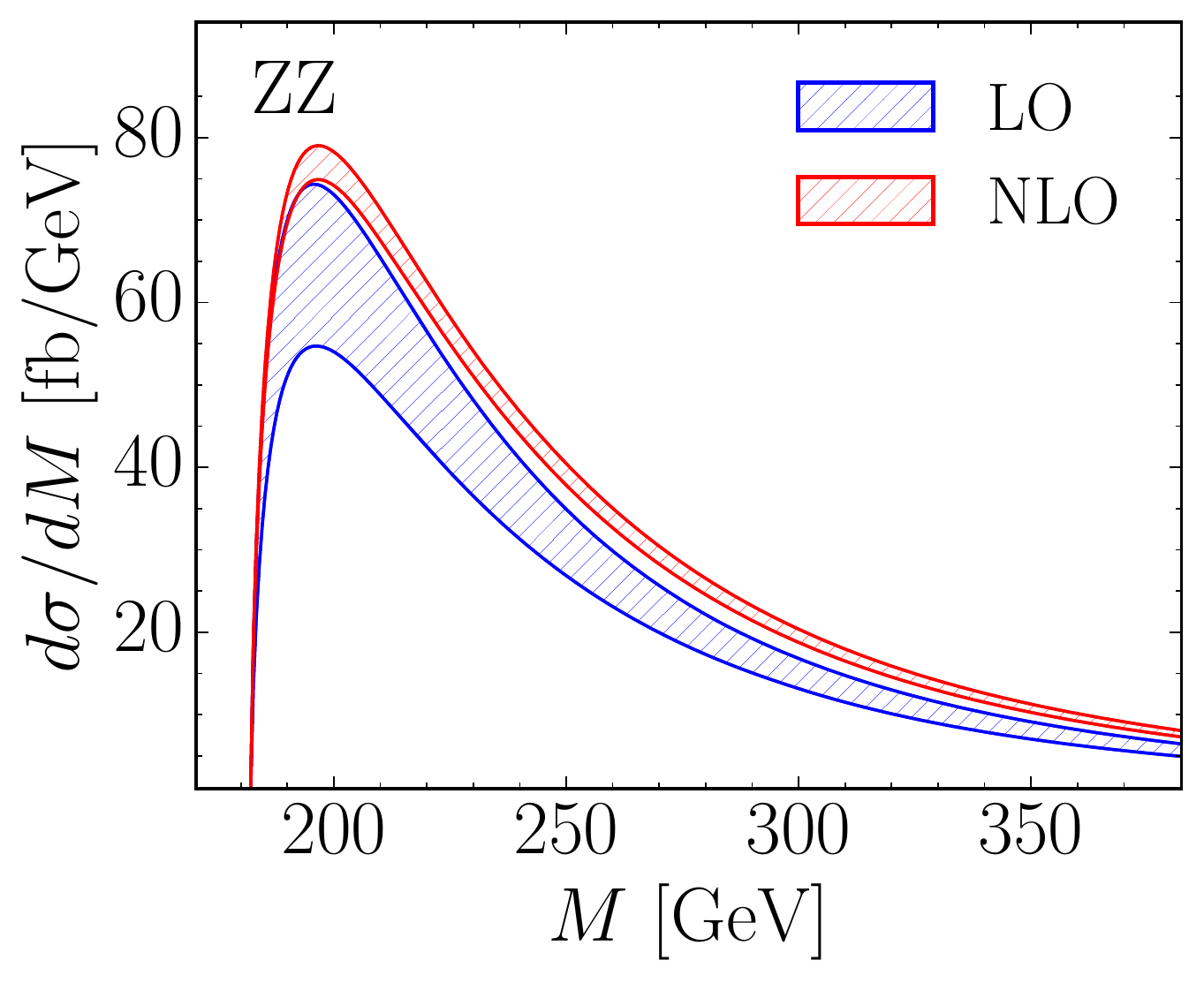}
        &
                \includegraphics[width=0.32\textwidth]{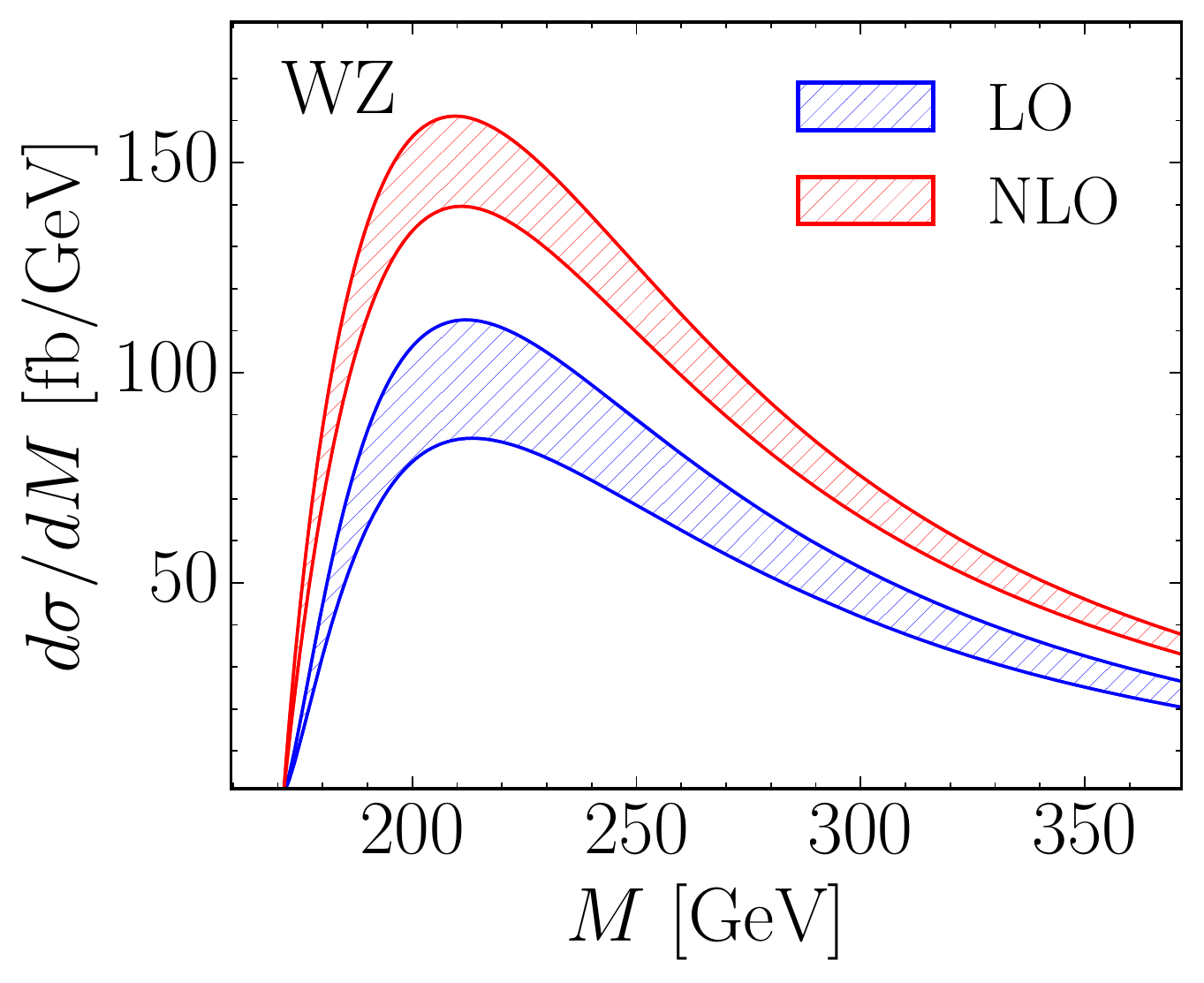}
	\\ \end{tabular}
        \caption{Differential cross-sections at LO (blue hatched bands) and NLO (red hatched band) are shown for 
        $W^+ W^-$ (left), $ZZ$ (center) and $W^\pm Z$ (right) production at $\sqrt{s}=8$ TeV LHC obtained from our 
        scale variation method in the complex $\mu^2$-plane as described in the text. Contributions from gluon-fusion
         channels are not included.}
        \label{f.8TeV_resum}
\end{figure*}

\subsection*{Gluon-induced Diboson Production}
$W^+ W^-$ and $ZZ$ production cross-sections get contributions from $gg$ channel, which although formally NNLO, 
can be sizable at the LHC owing to large gluon PDFs.  Although higher-order corrections to the process 
$gg \rightarrow VV'$ are currently unknown, they are expected to be large, as has been established for the case of 
Higgs production in the gluon-fusion channel where the NNLO K-factors can be as big as $\sim 2.5$.  In \cite{PiSq:Higgs},
it was shown that the bulk of such large K-factors stem from the $\pi^2$ enhanced terms. It is then reasonable to expect 
that the scale uncertainties in the theory predictions for $gg \rightarrow V V'$ process would be grossly underestimated
if the scale variation in the complex $\mu^2$-plane is not considered. The origin of large perturbative correction from 
complex-valued scales in gluon-induced diboson production as well as their resummation is identical to the arguments 
presented above for the $q \bar{q}'$ channel, which we briefly outline below.

The SCET Lagrangian for the $gg \rightarrow VV'$ process is similar to \eref{SCET:Lagrangian} with the SCET 
operator $\cj$ now describing the interaction of incoming `gluon-jets' with the outgoing electroweak gauge bosons.  
Analogous to \eref{SCET:current}, $\cj$ is constructed using gauge-invariant collinear and anti-collinear SCET gluon 
fields, $\ca^\mu_c$ and  $\ca^\mu_{\bar{c}}$. The hard coefficients analogous to \eref{HardCoeff} are only known at 
LO for the $gg \rightarrow VV'$ processes, however, their RG evolution is identical to that of the hard coefficient 
for the Higgs production, $gg \rightarrow h$. Therefore, it is possible to resum the large perturbative terms associated 
with the complex phase of $\mu^2$, so that a realistic estimate of the scale uncertainty can be obtained.  The evolution 
kernel $\cu (\upmu, \muh)$ required for the computation can be found in \cite{PiSq:Higgs}, which to a good 
approximation is given by
\beq
\begin{split}
\log \left| \cu (\upmu, \muh) \right|^2  &\approx \frac{\GA_0 \Theta^2 \alpha_s(\upmu)}{8 \pi} 
 \bigg[ 1 + \frac{\alpha_s(\upmu)}{4 \pi}   \bigg\{ \\
& \qquad  \frac{\GA_1}{\GA_0} -\beta_0 \log \left( \frac{M^2}{\upmu^2} \right) \bigg\} \bigg]
\end{split}
\label{e.kernel_gg}
\eeq  
where, $\GA_0=4 C_A$ and $\GA_1$ is given by the same expression as $\GF_1$ in \eref{GF1} but with $C_F$ 
replaced by $C_A$. 

To summarize this section, we have calculated the evolution kernels $\cu(\upmu, \muh)$, which resum the large 
perturbative corrections arising from the complex phase of the hard scale $\muh$.  In the next section, we will show
how to combine these kernels with the existing LO and NLO codes such as {\tt MCFM}, to estimate the true scale 
uncertainties and the central values for theory prediction of the diboson cross-sections.

\section{Results}
\label{s.results}

In order to incorporate our scale variation technique into existing fixed-order calculations, we require 
differential cross-sections in $M$.  Presently, only NLO $q \bar{q}' \rightarrow V V'$ and LO 
$g g \rightarrow V V'$ differential distributions are publicly available. All our numerical results will extensively 
use the {\tt MCFM} program for extracting the fixed-order differential cross-sections with {\tt MSTW2008} 
\cite{PDF:MSTW2008} as the choice for PDF sets. Implementing scale variation in the full complex 
$\mu^2$-plane requires computation of cross-section where the large perturbative corrections from the 
complex phase, $\Theta$ have been resummed. At LO, this is implemented as follows:
\beq
\frac{\der \sigma^{\text{LO}}}{\der M} (\Theta, \upmu, M)  = \left| \cu (\Theta, \upmu, M) \right|^2 
\frac{ \der \sigma^{\text{LO}}}{\der M}  (\upmu,M)
\eeq
where, $\cu(\Theta, \upmu, M) \equiv \cu(\upmu, \muh)$ is the evolution kernel defined in \eref{kernel} 
for the $q \bar{q}'$ channel and \eref{kernel_gg} for the $gg$ channel. The computation of kernels 
requires $\alpha_s(\upmu)$ which we obtain from NNLO PDF sets. The differential cross-section 
on the RHS of the above equation is obtained from {\tt MCFM} using LO PDF sets with the factorization scale 
$\muf = \upmu$. By varying the parameters $M/2 < \upmu < 2M$ and $-\pi < \Theta < 0$, we therefore obtain 
the scale variation in the complex $\mu^2$-plane.

Implementation of our method at NLO follows similarly, however, appropriate subtractions must be made to 
avoid double-counting since $\co(\alpha_s)$ contributions associated with the phase $\Theta$ are already 
a part of NLO. To do so, we remove $\co(\alpha_s)$ expansion of the kernels $\cu$ from the fixed-order 
NLO results, before implementing resummation. We therefore arrive at the following formula for the process
$q \bar{q}' \rightarrow VV'$ at NLO:     
\beq
\begin{split}
\frac{ \der \sigma^{\text{NLO}}}{\der M}(\Theta, \upmu, M)  &= \left| \cu (\Theta, \upmu, M) \right|^2 
\bigg[ \frac{\der \sigma^{\text{NLO}}}{\der M} (\upmu, M) \\
 &\qquad - \frac{\GF_0 \Theta^2 \alpha_s(\upmu)}{8 \pi} 
 \frac{\der \sigma^{\text{LO}}}{\der M}  (\upmu, M) \bigg]
\end{split}
\label{e.master}
\eeq
where, all quantities in the square bracket are computed using NLO PDF sets. The differential 
cross-sections on the RHS of the above equation are again obtained from {\tt MCFM} by setting the 
renormalization and factorization scales as $\mur = \muf = \upmu$. We are now in a position to present 
the numerical results for diboson production using our scale variation method.

\renewcommand{\arraystretch}{1.5}
\begin{table}[h]
\begin{ruledtabular}
%\begin{center}
\begin{tabular}{c|c|c|c|c|}
%\cline{2-5}
		& $7$ TeV			& $8$ TeV 		  	& $13$ TeV	 		& $14$ TeV 	 		\\ \hline \hline
\multicolumn{1} {r|}{$\sigma_{WW}^{\text{LO}}$ [pb]} 
		& $33.8 \pm 4.3$		& $41.2 \pm 5.5$		&$82.5 \pm 13.2$		&$91.4 \pm 15.0$		\\ \hline 
\multicolumn{1} {r|}{$\sigma^{\text{NLO}}_{WW}$ [pb]} 
		&$45.3 \pm 2.2$		& $55.2 \pm 2.6$		&$109.1 \pm 4.5$		&$120.7 \pm 5.0$		\\ \hline 
\multicolumn{1} {r|}{$\sigma^{gg}_{WW}$ [pb]} 
		&$1.6 \pm 0.7$			& $2.1 \pm 1.0$		&$6.0 \pm 2.7$			&$7.0 \pm 3.1$			 
\end{tabular}
%\end{center}
\end{ruledtabular}
\caption{LO and NLO cross-section predictions for $W^+ W^-$ production at $\sqrt{s}=7,8,13$ and $14$ TeV LHC runs, 
using our scale-variation method. The contribution of gluon-fusion channel is shown separately. } 
\label{t.WW} 
\end{table}
\begin{table}[h]
\centering
\begin{ruledtabular}
\begin{tabular}{c|c|c|c|c|}
%\cline{2-5}
		& $7$ TeV			& $8$ TeV 		  	& $13$ TeV	 		& $14$ TeV 	 		\\ \hline \hline
\multicolumn{1} {r|}{$\sigma_{ZZ}^{\text{LO}}$ [pb]} 
		& $4.7 \pm 0.6$		& $5.8 \pm 0.7$		&$11.9 \pm 1.8$		&$13.2 \pm 2.1$		\\ \hline 
\multicolumn{1} {r|}{$\sigma^{\text{NLO}}_{ZZ}$ [pb]} 
		&$6.0 \pm 0.2$			& $7.3 \pm 0.2$		&$14.6 \pm 0.4$		&$16.2 \pm 0.4$		\\ \hline 
\multicolumn{1} {r|}{$\sigma^{gg}_{ZZ}$ [pb]} 
		&$0.5 \pm 0.2$			& $0.7 \pm 0.3$		&$1.9 \pm 0.9$			&$2.2 \pm 1.0$			 
\end{tabular}
\caption{Same as \tref{WW} but for $ZZ$ production.}
\label{t.ZZ}
\end{ruledtabular}
\end{table}
\begin{table}[h]
\begin{ruledtabular}
\centering
\begin{tabular}{c|c|c|c|c|}
%\cline{2-5}
		& $7$ TeV			& $8$ TeV 		  	& $13$ TeV	 		& $14$ TeV 	 		\\ \hline \hline
\multicolumn{1} {r|}{$\sigma_{W^+Z}^{\text{LO}}$ [pb]} 
		& $7.8 \pm 1.0$		& $9.4 \pm 1.1$		&$18.4 \pm 2.6$		&$20.3 \pm 3.0$		\\ \hline 
\multicolumn{1} {r|}{$\sigma^{\text{NLO}}_{W^+Z}$ [pb]} 
		&$11.6 \pm 0.8$		& $14.2 \pm 1.0$		&$28.3 \pm 1.9$		&$31.6 \pm 2.3$		\\ \hline 
\multicolumn{1} {r|}{$\sigma^{\text{LO}}_{W^-Z}$ [pb]} 
		&$4.2 \pm 0.5$			& $5.3 \pm 0.6$		&$11.3 \pm 1.6$		&$12.7 \pm 1.9$		\\ \hline 
\multicolumn{1} {r|}{$\sigma^{\text{NLO}}_{W^-Z}$ [pb]} 
		&$6.5 \pm 0.5$			& $8.2 \pm 0.6$		&$18.3 \pm 1.4$		&$20.3 \pm 1.4$		 
\end{tabular}
\caption{Same as \tref{WW} but for $W^\pm Z$ production. There is no gluon-fusion production channel for this process.}
\label{t.WZ}
\end{ruledtabular}
\end{table}

The total LO and NLO cross-sections, along with their scale uncertainties, for $W^+ W^-$, $ZZ$ and $W^\pm Z$  
production at different center of mass energy LHC runs are presented in \tref{WW}, \tref{ZZ} and \tref{WZ}, respectively. 
The gluon-fusion contribution for  $W^+ W^-$ and $ZZ$ processes  is also shown in these tables. 
Besides the scale uncertainties shown in the tables, there are additional theoretical uncertainties of $\sim 3$--$4\%$ 
from the PDFs.
\begin{figure*}
        \center
        \begin{tabular}{ccc}
                \includegraphics[width=0.32\textwidth]{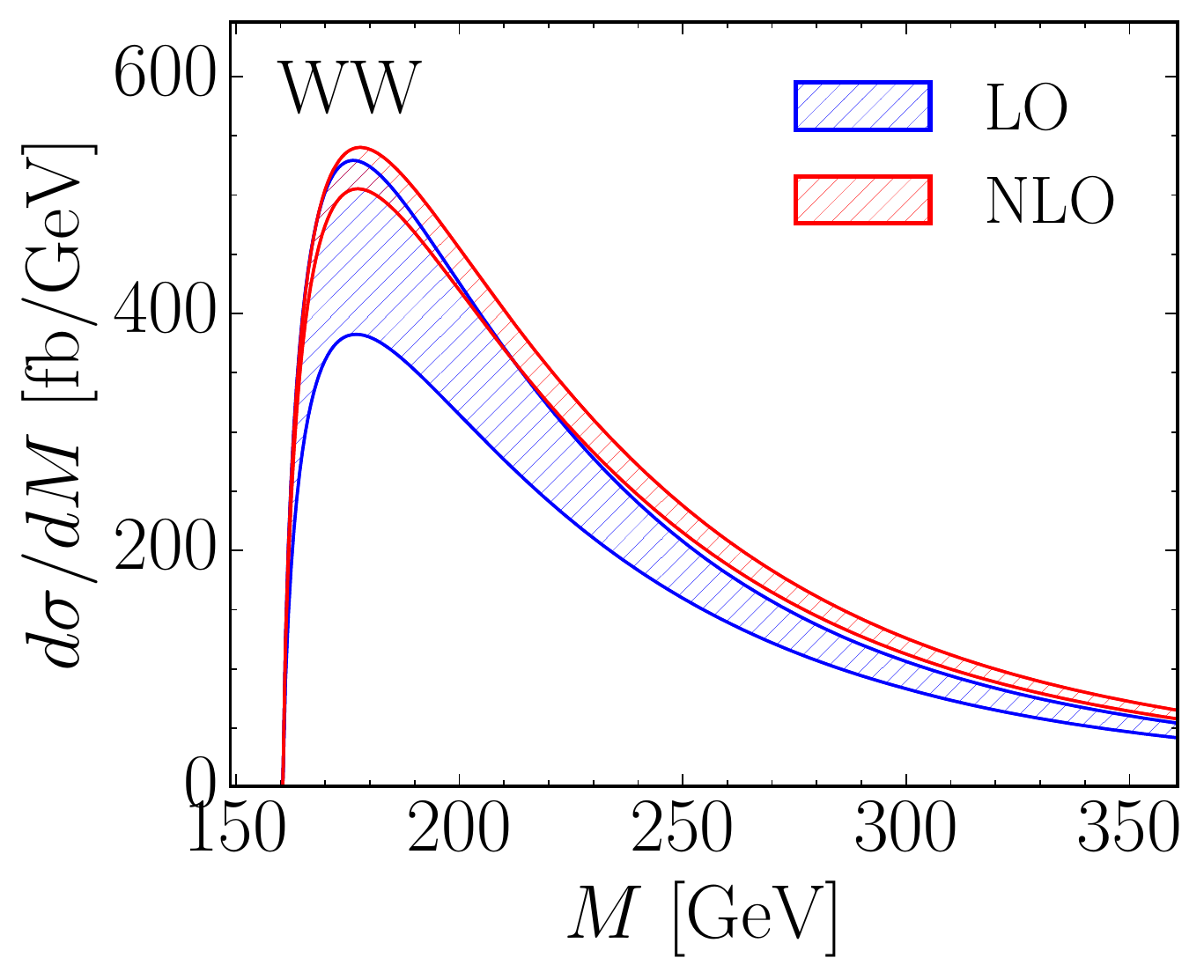}
        &
                \includegraphics[width=0.32\textwidth]{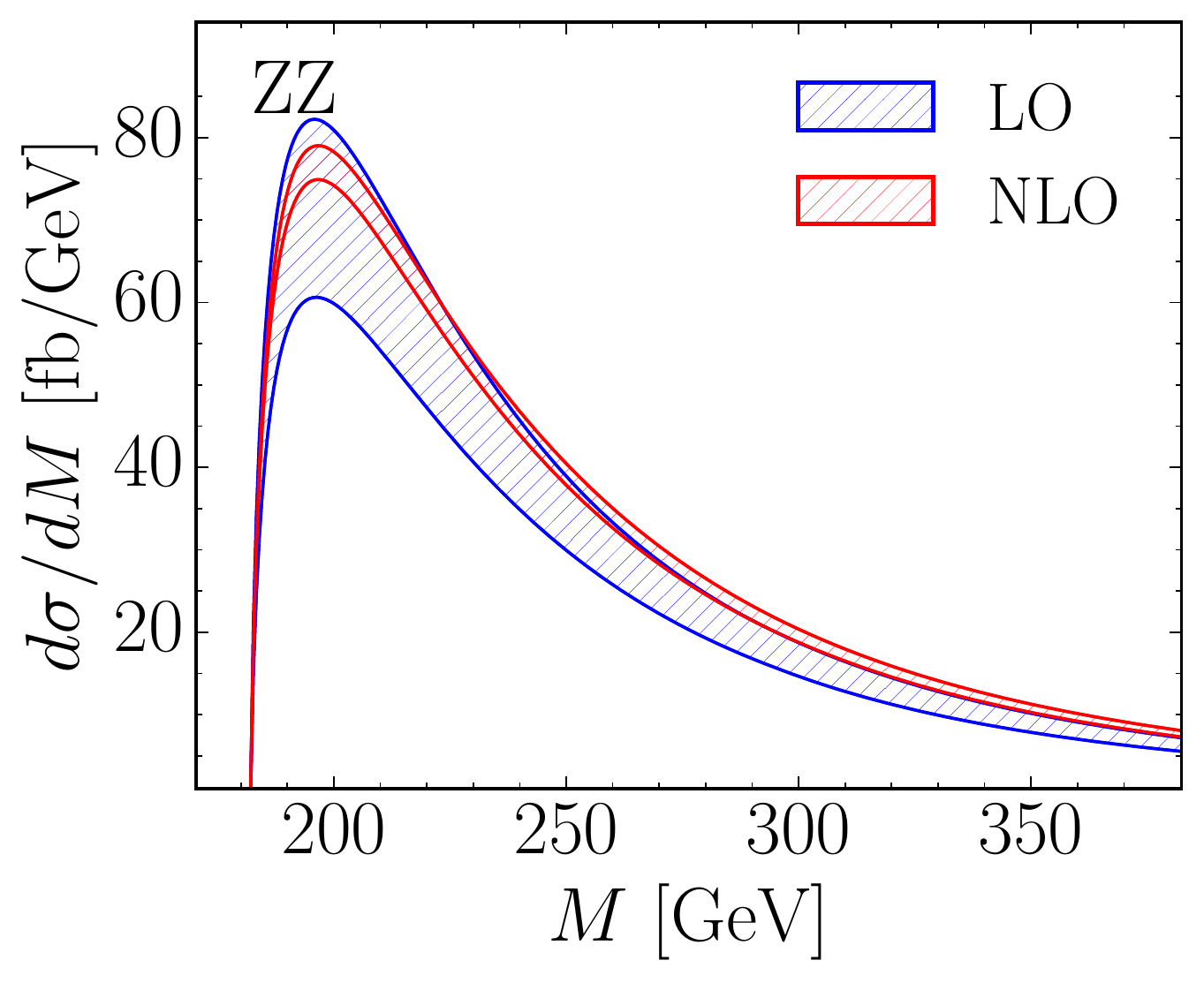}
        &
                \includegraphics[width=0.32\textwidth]{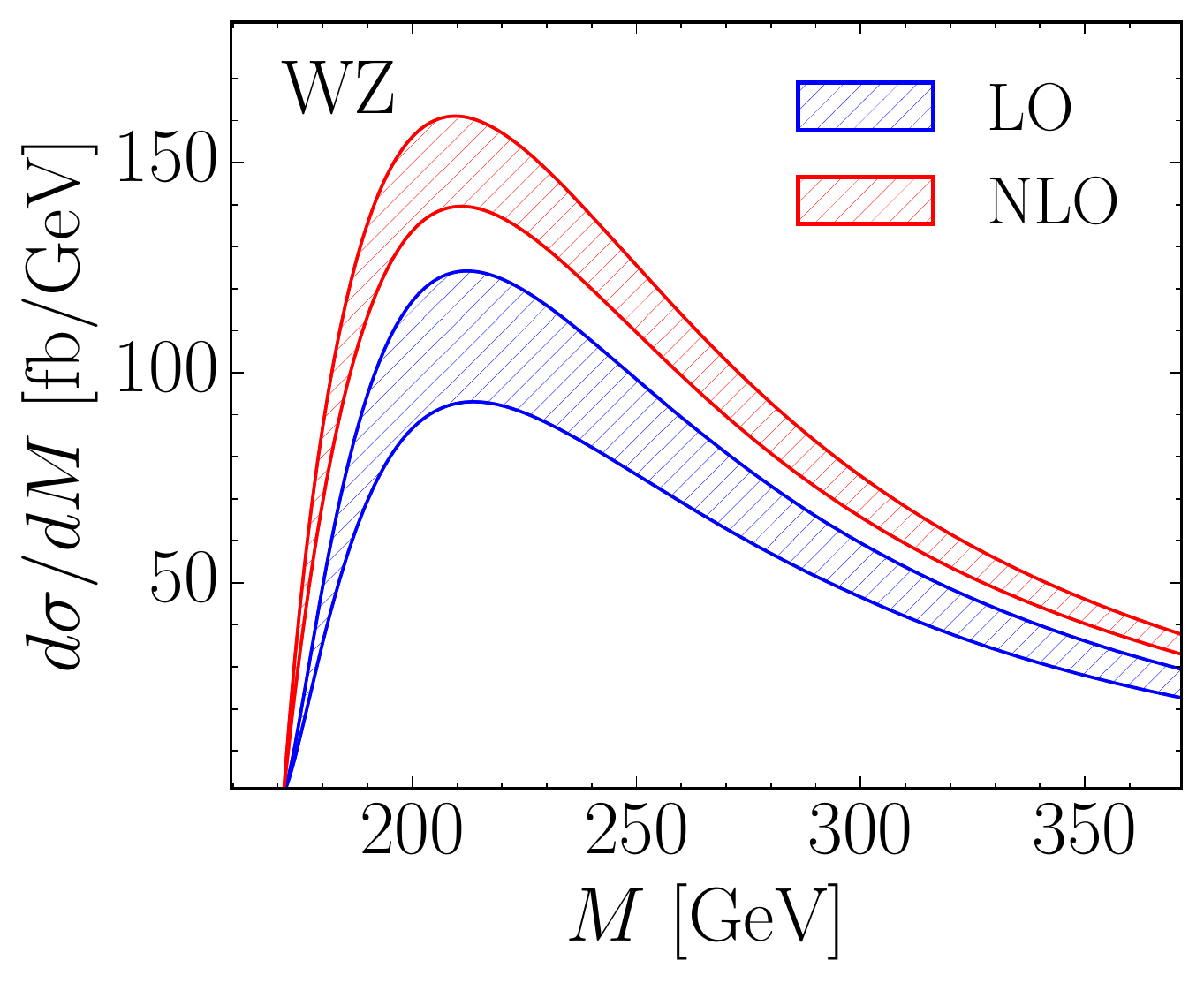}
	\\ \end{tabular}
        \caption{Same as \fref{8TeV_resum} but NLO PDF set is used for both LO and NLO cross-sections.}
        \label{f.8TeV_PDF}
\end{figure*}
In \fref{8TeV_resum}, LO and NLO differential cross-sections are shown for $W^+ W^-$, $ZZ$ and $W^+ Z$ 
production in the $q \bar{q}'$ channel at $\sqrt{s}=8$ TeV LHC run using the complex scale variation technique 
described earlier in this section. This is to be contrasted with \fref{8TeV_NLO} where traditional 
approach ($M/2 < \mu < 2M$) for estimating the scale uncertainties was followed.  We conclude with the 
following remarks: 
\begin{itemize}
\item
First and foremost, we find that the scale uncertainties are grossly underestimated in \fref{8TeV_NLO}, 
corroborating our argument that variation in the full complex $\mu^2$-plane must be considered in order
to estimate the true scale uncertainties. As can be seen in \fref{8TeV_resum}, the effect is striking
at LO, with uncertainties in the range $13$--$16\%$  in contrast to the traditional approach  which estimate the 
uncertainties to be $2$--$4\%$.  This is particularly relevant for diboson production in gluon-fusion channels which,
although formally NNLO, is absent at lower orders, thus suffering from the same 
scale uncertainty underestimation issues as LO $q \bar{q}'$ channel.  Even at NLO, the scale uncertainties using our 
method are $3$--$4\%$ higher compared to previous fixed-order estimates with $M/2 < \mu < 2M$. Most importantly,
NLO predictions for all diboson processes are now less than $5\sigma$ away from their LO value, giving us confidence 
in the reliability of the perturbation theory. 
\item
Owing to the large scale uncertainties in our method of scale variation, the central value predictions for LO 
and NLO cross-sections are also altered. We assign the central value to be the center of the uncertainty band so 
that all the numerical results presented in this section have symmetrical error bars. Our best prediction 
for LO cross-sections are $15\%$ ($46\%$) higher than the previous fixed-order predictions for 
$q \bar{q}'$ ($gg$) channel. At NLO, we find a marginal increase of $3$--$4\%$ for the $q \bar{q}'$ channel. 
\item
Up to this point, we have compared our results with fixed-order calculations that use dynamic scales, 
$M/2 < \mu < 2M$. We now compare our results with fixed-order predictions that use a rather ad-hoc choice for
the renormalization and factorization scales $\overline{m}/2 < \mu < 2\overline{m}$, where $\overline{m}$ is the 
average mass of the vector bosons $V$ and $V'$ \cite{VV:NLOfull+gg}. At LO, the fixed-order calculations fail 
miserably as expected, predicting scale uncertainty as low as $1\%$. However, at NLO, the fixed-order 
calculations work surprisingly well with their central-value predictions within $2\%$ of our results for all diboson 
processes. The scale uncertainties in these calculations are marginally underestimated by $2$--$4\%$ with the 
largest discrepancy with our method arising for the $W^\pm Z$ process. In general, we expect that our method 
assigns larger scale uncertainty to processes with large K-factor.   
\item
While we have consistently used LO PDF set for LO cross-sections and NLO PDF set for NLO cross-sections, it is 
worthwhile to gauge the impact of the order of PDF on higher-order corrections. In \fref{8TeV_PDF}, we present 
our results using NLO PDF set for both LO and NLO computations. Clearly, a significant portion of higher order 
corrections to diboson processes is driven by the radiative corrections to the PDF. 
\item
As we already discussed, closely related calculations have appeared in the literature before under the name of 
$\pi^2$-resummation, where the complex phase is held fixed at $\Theta = - \pi + 0^+$. $\pi^2$-resummation 
for $ZZ$ and $W^\pm Z$ channels was performed in \cite{VZ:Threshold} leading to an increase over fixed-order 
NLO predictions by $4 \%$ and $8\%$ respectively. Given that we vary $-\pi < \Theta < 0$, such an enhancement 
appears as an upper limit of our scale variation, with an increase in our central value prediction being nearly half
of that from $\pi^2$-resummation. Similar calculations for $W^+ W^-$ production in the $0$-jet bin \cite{WW:JetVeto}
reveal impact of $\pi^2$-resummation resummation to be $\sim 7\%$ beyond NLO, thus partly explaining the 
slight excess in the $W^+W^-$ cross-section measurement compared to the theory prediction\footnote{The other 
factor possibly responsible for the discrepancy between the experiment and the theory predictions of $W^+W^-$ 
cross-section is the jet-veto efficiency which has been explored in  \cite{WW:JetVeto, WW:pT2, WW:JetVeto2}.}.  
\item 
Recently, NNLO calculations have been performed for $W^+ W^-$ \cite{WW:NNLO} and $ZZ$ \cite{ZZ:NNLO} 
processes, although differential distributions in $M$ are not publicly available  so that our scale variation 
technique can not be applied at the moment. It would nevertheless be an interesting check if the choice of 
scale considered in these calculations, $\overline{m}/2 < \mu < 2\overline{m}$, can mimic our method even at 
NNLO.  It should be noted that the increase beyond NLO from $\pi^2$-resummation discussed above is 
comparable to that from full NNLO calculations for both $W^+W^-$ and $ZZ$ 
channels suggesting that $\pi^2$ terms dominate even beyond NLO.  We have already pointed out that 
there are production channels, which open up only at higher orders, for which scale uncertainties would be 
underestimated. One such channel is the gluon-fusion mode at NNLO for which we have explicitly presented
the central value and the scale uncertainties.   
\item
Finally, we point out that our scale-variation technique can be easily incorporated when considering other higher 
order calculations to diboson processes by simply modifying \eref{master}. For fixed order calculations, one simply
 has to replace $\der \sigma / \der M$, while for resummation calculations, the evolution kernel $\cu$ also gets modified. 
In particular, there are two scenarios where large perturbative corrections from complex-phase of the scales
is partially cancelled at large $M$. As shown in \cite{WW:EW1, WW:EW2, VV:EW1, VV:EW2}, for large $M$, 
electroweak Sudakov logarithms  of the form $\alpha \log^2(M_V/M)$ and $\alpha \log(M_V/M)$ can lead to 
large negative corrections to diboson cross-sections. Another scenario where cancellation at large $M$ is realized
 is the presence of jet-vetoes, where logarithms of the form $\alpha_s \log^2(\ptv/M)$  and $\alpha_s \log (\ptv/M)$ 
 again lead to large negative corrections \cite{WW:JetVeto}.  Further, our method of scale variation can be easily 
 extended to beyond NLO calculations as well as to other LHC processes, involving more complicated colored 
 structures in the final states. 
\end{itemize}

%\kern-3em       
\subsection*{Acknowledgements}
The author would like to thank Sally Dawson, Patrick Meade, Takemichi Okui and George Sterman for useful 
discussions and comments on the manuscript.

%%%%%%%%%%%%%%%%%%%%%%%%%%%%%%%%%%%%%%%%%%%%%%%%%%%%%%%%%%%%%%%%%%%%%%%%%%%%
%%%%%%%%%%%%%%%%%%%%%%%%%%%%%%%%%%%%%%%%%%%%%%%%%%%%%%%%%%%%%%%%%%%%%%%%%%%%

\bibliographystyle{apsrev4-1}
\bibliography{refs}

\begin{thebibliography}{71}%
\makeatletter
\providecommand \@ifxundefined [1]{%
 \@ifx{#1\undefined}
}%
\providecommand \@ifnum [1]{%
 \ifnum #1\expandafter \@firstoftwo
 \else \expandafter \@secondoftwo
 \fi
}%
\providecommand \@ifx [1]{%
 \ifx #1\expandafter \@firstoftwo
 \else \expandafter \@secondoftwo
 \fi
}%
\providecommand \natexlab [1]{#1}%
\providecommand \enquote  [1]{``#1''}%
\providecommand \bibnamefont  [1]{#1}%
\providecommand \bibfnamefont [1]{#1}%
\providecommand \citenamefont [1]{#1}%
\providecommand \href@noop [0]{\@secondoftwo}%
\providecommand \href [0]{\begingroup \@sanitize@url \@href}%
\providecommand \@href[1]{\@@startlink{#1}\@@href}%
\providecommand \@@href[1]{\endgroup#1\@@endlink}%
\providecommand \@sanitize@url [0]{\catcode `\\12\catcode `\$12\catcode
  `\&12\catcode `\#12\catcode `\^12\catcode `\_12\catcode `\%12\relax}%
\providecommand \@@startlink[1]{}%
\providecommand \@@endlink[0]{}%
\providecommand \url  [0]{\begingroup\@sanitize@url \@url }%
\providecommand \@url [1]{\endgroup\@href {#1}{\urlprefix }}%
\providecommand \urlprefix  [0]{URL }%
\providecommand \Eprint [0]{\href }%
\providecommand \doibase [0]{http://dx.doi.org/}%
\providecommand \selectlanguage [0]{\@gobble}%
\providecommand \bibinfo  [0]{\@secondoftwo}%
\providecommand \bibfield  [0]{\@secondoftwo}%
\providecommand \translation [1]{[#1]}%
\providecommand \BibitemOpen [0]{}%
\providecommand \bibitemStop [0]{}%
\providecommand \bibitemNoStop [0]{.\EOS\space}%
\providecommand \EOS [0]{\spacefactor3000\relax}%
\providecommand \BibitemShut  [1]{\csname bibitem#1\endcsname}%
\let\auto@bib@innerbib\@empty
%</preamble>
\bibitem [{ATL(2014)}]{ATLAS:WW8}%
  \BibitemOpen
  \href@noop {} {\emph {\bibinfo {title} {{Measurement of the $W^+W^-$
  production cross section in proton-proton collisions at $\sqrt{s} =8$ TeV
  with the ATLAS detector}}}},\ \bibinfo {type} {Tech. Rep.}\ \bibinfo {number}
  {ATLAS-CONF-2014-033}\ (\bibinfo  {institution} {CERN},\ \bibinfo {address}
  {Geneva},\ \bibinfo {year} {2014})\BibitemShut {NoStop}%
\bibitem [{ATL(2013{\natexlab{a}})}]{ATLAS:ZZ8}%
  \BibitemOpen
  \href@noop {} {\emph {\bibinfo {title} {{Measurement of the total ZZ
  production cross section in proton-proton collisions at $\sqrt{s} = 8$ TeV in
  20 fb$^{-1}$ with the ATLAS detector}}}},\ \bibinfo {type} {Tech. Rep.}\
  \bibinfo {number} {ATLAS-CONF-2013-020}\ (\bibinfo  {institution} {CERN},\
  \bibinfo {address} {Geneva},\ \bibinfo {year} {2013})\BibitemShut {NoStop}%
\bibitem [{ATL(2013{\natexlab{b}})}]{ATLAS:WZ8}%
  \BibitemOpen
  \href@noop {} {\emph {\bibinfo {title} {{A Measurement of WZ Production in
  Proton-Proton Collisions at sqrt(s)=8TeV with the ATLAS Detector}}}},\
  \bibinfo {type} {Tech. Rep.}\ \bibinfo {number} {ATLAS-CONF-2013-021}\
  (\bibinfo  {institution} {CERN},\ \bibinfo {address} {Geneva},\ \bibinfo
  {year} {2013})\BibitemShut {NoStop}%
\bibitem [{\citenamefont {Aad}\ \emph {et~al.}(2013{\natexlab{a}})\citenamefont
  {Aad} \emph {et~al.}}]{ATLAS:WW7}%
  \BibitemOpen
  \bibfield  {author} {\bibinfo {author} {\bibfnamefont {G.}~\bibnamefont
  {Aad}} \emph {et~al.} (\bibinfo {collaboration} {ATLAS}),\ }\href {\doibase
  10.1103/PhysRevD.87.112001, 10.1103/PhysRevD.88.079906} {\bibfield  {journal}
  {\bibinfo  {journal} {Phys.Rev.}\ }\textbf {\bibinfo {volume} {D87}},\
  \bibinfo {pages} {112001} (\bibinfo {year} {2013}{\natexlab{a}})},\ \Eprint
  {http://arxiv.org/abs/1210.2979} {arXiv:1210.2979 [hep-ex]} \BibitemShut
  {NoStop}%
%%CITATION = ARXIV:1210.2979;%%
\bibitem [{\citenamefont {Aad}\ \emph {et~al.}(2013{\natexlab{b}})\citenamefont
  {Aad} \emph {et~al.}}]{ATLAS:ZZ7}%
  \BibitemOpen
  \bibfield  {author} {\bibinfo {author} {\bibfnamefont {G.}~\bibnamefont
  {Aad}} \emph {et~al.} (\bibinfo {collaboration} {ATLAS}),\ }\href {\doibase
  10.1007/JHEP03(2013)128} {\bibfield  {journal} {\bibinfo  {journal} {JHEP}\
  }\textbf {\bibinfo {volume} {1303}},\ \bibinfo {pages} {128} (\bibinfo {year}
  {2013}{\natexlab{b}})},\ \Eprint {http://arxiv.org/abs/1211.6096}
  {arXiv:1211.6096 [hep-ex]} \BibitemShut {NoStop}%
%%CITATION = ARXIV:1211.6096;%%
\bibitem [{\citenamefont {Aad}\ \emph {et~al.}(2012)\citenamefont {Aad} \emph
  {et~al.}}]{ATLAS:WZ7}%
  \BibitemOpen
  \bibfield  {author} {\bibinfo {author} {\bibfnamefont {G.}~\bibnamefont
  {Aad}} \emph {et~al.} (\bibinfo {collaboration} {ATLAS}),\ }\href {\doibase
  10.1140/epjc/s10052-012-2173-0} {\bibfield  {journal} {\bibinfo  {journal}
  {Eur.Phys.J.}\ }\textbf {\bibinfo {volume} {C72}},\ \bibinfo {pages} {2173}
  (\bibinfo {year} {2012})},\ \Eprint {http://arxiv.org/abs/1208.1390}
  {arXiv:1208.1390 [hep-ex]} \BibitemShut {NoStop}%
%%CITATION = ARXIV:1208.1390;%%
\bibitem [{\citenamefont {Chatrchyan}\ \emph
  {et~al.}(2013{\natexlab{a}})\citenamefont {Chatrchyan} \emph
  {et~al.}}]{CMS:WW8}%
  \BibitemOpen
  \bibfield  {author} {\bibinfo {author} {\bibfnamefont {S.}~\bibnamefont
  {Chatrchyan}} \emph {et~al.} (\bibinfo {collaboration} {CMS Collaboration}),\
  }\href {\doibase 10.1016/j.physletb.2013.03.027} {\bibfield  {journal}
  {\bibinfo  {journal} {Phys.Lett.}\ }\textbf {\bibinfo {volume} {B721}},\
  \bibinfo {pages} {190} (\bibinfo {year} {2013}{\natexlab{a}})},\ \Eprint
  {http://arxiv.org/abs/1301.4698} {arXiv:1301.4698 [hep-ex]} \BibitemShut
  {NoStop}%
%%CITATION = ARXIV:1301.4698;%%
\bibitem [{\citenamefont {Khachatryan}\ \emph {et~al.}(2014)\citenamefont
  {Khachatryan} \emph {et~al.}}]{CMS:ZZ8}%
  \BibitemOpen
  \bibfield  {author} {\bibinfo {author} {\bibfnamefont {V.}~\bibnamefont
  {Khachatryan}} \emph {et~al.} (\bibinfo {collaboration} {CMS
  Collaboration}),\ }\href@noop {} {\  (\bibinfo {year} {2014})},\ \Eprint
  {http://arxiv.org/abs/1406.0113} {arXiv:1406.0113 [hep-ex]} \BibitemShut
  {NoStop}%
%%CITATION = ARXIV:1406.0113;%%
\bibitem [{CMS(2013)}]{CMS:WZ7and8}%
  \BibitemOpen
  \href@noop {} {\emph {\bibinfo {title} {{Measurement of WZ production
  rate}}}},\ \bibinfo {type} {Tech. Rep.}\ \bibinfo {number}
  {CMS-PAS-SMP-12-006}\ (\bibinfo  {institution} {CERN},\ \bibinfo {address}
  {Geneva},\ \bibinfo {year} {2013})\BibitemShut {NoStop}%
\bibitem [{\citenamefont {Chatrchyan}\ \emph
  {et~al.}(2013{\natexlab{b}})\citenamefont {Chatrchyan} \emph
  {et~al.}}]{CMS:WW7}%
  \BibitemOpen
  \bibfield  {author} {\bibinfo {author} {\bibfnamefont {S.}~\bibnamefont
  {Chatrchyan}} \emph {et~al.} (\bibinfo {collaboration} {CMS Collaboration}),\
  }\href {\doibase 10.1140/epjc/s10052-013-2610-8} {\bibfield  {journal}
  {\bibinfo  {journal} {Eur.Phys.J.}\ }\textbf {\bibinfo {volume} {C73}},\
  \bibinfo {pages} {2610} (\bibinfo {year} {2013}{\natexlab{b}})},\ \Eprint
  {http://arxiv.org/abs/1306.1126} {arXiv:1306.1126 [hep-ex]} \BibitemShut
  {NoStop}%
%%CITATION = ARXIV:1306.1126;%%
\bibitem [{\citenamefont {Chatrchyan}\ \emph
  {et~al.}(2013{\natexlab{c}})\citenamefont {Chatrchyan} \emph
  {et~al.}}]{CMS:ZZ7}%
  \BibitemOpen
  \bibfield  {author} {\bibinfo {author} {\bibfnamefont {S.}~\bibnamefont
  {Chatrchyan}} \emph {et~al.} (\bibinfo {collaboration} {CMS Collaboration}),\
  }\href {\doibase 10.1007/JHEP01(2013)063} {\bibfield  {journal} {\bibinfo
  {journal} {JHEP}\ }\textbf {\bibinfo {volume} {1301}},\ \bibinfo {pages}
  {063} (\bibinfo {year} {2013}{\natexlab{c}})},\ \Eprint
  {http://arxiv.org/abs/1211.4890} {arXiv:1211.4890 [hep-ex]} \BibitemShut
  {NoStop}%
%%CITATION = ARXIV:1211.4890;%%
\bibitem [{\citenamefont {Feigl}\ \emph {et~al.}(2012)\citenamefont {Feigl},
  \citenamefont {Rzehak},\ and\ \citenamefont {Zeppenfeld}}]{NewPhyWW:1}%
  \BibitemOpen
  \bibfield  {author} {\bibinfo {author} {\bibfnamefont {B.}~\bibnamefont
  {Feigl}}, \bibinfo {author} {\bibfnamefont {H.}~\bibnamefont {Rzehak}}, \
  and\ \bibinfo {author} {\bibfnamefont {D.}~\bibnamefont {Zeppenfeld}},\
  }\href {\doibase 10.1016/j.physletb.2012.09.033} {\bibfield  {journal}
  {\bibinfo  {journal} {Phys.Lett.}\ }\textbf {\bibinfo {volume} {B717}},\
  \bibinfo {pages} {390} (\bibinfo {year} {2012})},\ \Eprint
  {http://arxiv.org/abs/1205.3468} {arXiv:1205.3468 [hep-ph]} \BibitemShut
  {NoStop}%
%%CITATION = ARXIV:1205.3468;%%
\bibitem [{\citenamefont {Curtin}\ \emph
  {et~al.}(2013{\natexlab{a}})\citenamefont {Curtin}, \citenamefont {Jaiswal},\
  and\ \citenamefont {Meade}}]{NewPhyWW:2}%
  \BibitemOpen
  \bibfield  {author} {\bibinfo {author} {\bibfnamefont {D.}~\bibnamefont
  {Curtin}}, \bibinfo {author} {\bibfnamefont {P.}~\bibnamefont {Jaiswal}}, \
  and\ \bibinfo {author} {\bibfnamefont {P.}~\bibnamefont {Meade}},\ }\href
  {\doibase 10.1103/PhysRevD.87.031701} {\bibfield  {journal} {\bibinfo
  {journal} {Phys.Rev.}\ }\textbf {\bibinfo {volume} {D87}},\ \bibinfo {pages}
  {031701} (\bibinfo {year} {2013}{\natexlab{a}})},\ \Eprint
  {http://arxiv.org/abs/1206.6888} {arXiv:1206.6888 [hep-ph]} \BibitemShut
  {NoStop}%
%%CITATION = ARXIV:1206.6888;%%
\bibitem [{\citenamefont {Jaiswal}\ \emph {et~al.}(2013)\citenamefont
  {Jaiswal}, \citenamefont {Kopp},\ and\ \citenamefont {Okui}}]{NewPhyWW:3}%
  \BibitemOpen
  \bibfield  {author} {\bibinfo {author} {\bibfnamefont {P.}~\bibnamefont
  {Jaiswal}}, \bibinfo {author} {\bibfnamefont {K.}~\bibnamefont {Kopp}}, \
  and\ \bibinfo {author} {\bibfnamefont {T.}~\bibnamefont {Okui}},\ }\href
  {\doibase 10.1103/PhysRevD.87.115017} {\bibfield  {journal} {\bibinfo
  {journal} {Phys.Rev.}\ }\textbf {\bibinfo {volume} {D87}},\ \bibinfo {pages}
  {115017} (\bibinfo {year} {2013})},\ \Eprint {http://arxiv.org/abs/1303.1181}
  {arXiv:1303.1181 [hep-ph]} \BibitemShut {NoStop}%
%%CITATION = ARXIV:1303.1181;%%
\bibitem [{\citenamefont {Rolbiecki}\ and\ \citenamefont
  {Sakurai}(2013)}]{NewPhyWW:4}%
  \BibitemOpen
  \bibfield  {author} {\bibinfo {author} {\bibfnamefont {K.}~\bibnamefont
  {Rolbiecki}}\ and\ \bibinfo {author} {\bibfnamefont {K.}~\bibnamefont
  {Sakurai}},\ }\href {\doibase 10.1007/JHEP09(2013)004} {\bibfield  {journal}
  {\bibinfo  {journal} {JHEP}\ }\textbf {\bibinfo {volume} {1309}},\ \bibinfo
  {pages} {004} (\bibinfo {year} {2013})},\ \Eprint
  {http://arxiv.org/abs/1303.5696} {arXiv:1303.5696 [hep-ph]} \BibitemShut
  {NoStop}%
%%CITATION = ARXIV:1303.5696;%%
\bibitem [{\citenamefont {Curtin}\ \emph
  {et~al.}(2013{\natexlab{b}})\citenamefont {Curtin}, \citenamefont {Jaiswal},
  \citenamefont {Meade},\ and\ \citenamefont {Tien}}]{NewPhyWW:5}%
  \BibitemOpen
  \bibfield  {author} {\bibinfo {author} {\bibfnamefont {D.}~\bibnamefont
  {Curtin}}, \bibinfo {author} {\bibfnamefont {P.}~\bibnamefont {Jaiswal}},
  \bibinfo {author} {\bibfnamefont {P.}~\bibnamefont {Meade}}, \ and\ \bibinfo
  {author} {\bibfnamefont {P.-J.}\ \bibnamefont {Tien}},\ }\href {\doibase
  10.1007/JHEP08(2013)068} {\bibfield  {journal} {\bibinfo  {journal} {JHEP}\
  }\textbf {\bibinfo {volume} {1308}},\ \bibinfo {pages} {068} (\bibinfo {year}
  {2013}{\natexlab{b}})},\ \Eprint {http://arxiv.org/abs/1304.7011}
  {arXiv:1304.7011 [hep-ph]} \BibitemShut {NoStop}%
%%CITATION = ARXIV:1304.7011;%%
\bibitem [{\citenamefont {Curtin}\ \emph {et~al.}(2014)\citenamefont {Curtin},
  \citenamefont {Meade},\ and\ \citenamefont {Tien}}]{NewPhyWW:6}%
  \BibitemOpen
  \bibfield  {author} {\bibinfo {author} {\bibfnamefont {D.}~\bibnamefont
  {Curtin}}, \bibinfo {author} {\bibfnamefont {P.}~\bibnamefont {Meade}}, \
  and\ \bibinfo {author} {\bibfnamefont {P.-J.}\ \bibnamefont {Tien}},\
  }\href@noop {} {\  (\bibinfo {year} {2014})},\ \Eprint
  {http://arxiv.org/abs/1406.0848} {arXiv:1406.0848 [hep-ph]} \BibitemShut
  {NoStop}%
%%CITATION = ARXIV:1406.0848;%%
\bibitem [{\citenamefont {Kim}\ \emph {et~al.}(2014)\citenamefont {Kim},
  \citenamefont {Rolbiecki}, \citenamefont {Sakurai},\ and\ \citenamefont
  {Tattersall}}]{NewPhyWW:7}%
  \BibitemOpen
  \bibfield  {author} {\bibinfo {author} {\bibfnamefont {J.~S.}\ \bibnamefont
  {Kim}}, \bibinfo {author} {\bibfnamefont {K.}~\bibnamefont {Rolbiecki}},
  \bibinfo {author} {\bibfnamefont {K.}~\bibnamefont {Sakurai}}, \ and\
  \bibinfo {author} {\bibfnamefont {J.}~\bibnamefont {Tattersall}},\
  }\href@noop {} {\  (\bibinfo {year} {2014})},\ \Eprint
  {http://arxiv.org/abs/1406.0858} {arXiv:1406.0858 [hep-ph]} \BibitemShut
  {NoStop}%
%%CITATION = ARXIV:1406.0858;%%
\bibitem [{\citenamefont {Luo}\ \emph {et~al.}(2014)\citenamefont {Luo},
  \citenamefont {Luo}, \citenamefont {Wang}, \citenamefont {Xu},\ and\
  \citenamefont {Zhu}}]{NewPhyWW:8}%
  \BibitemOpen
  \bibfield  {author} {\bibinfo {author} {\bibfnamefont {H.}~\bibnamefont
  {Luo}}, \bibinfo {author} {\bibfnamefont {M.-x.}\ \bibnamefont {Luo}},
  \bibinfo {author} {\bibfnamefont {K.}~\bibnamefont {Wang}}, \bibinfo {author}
  {\bibfnamefont {T.}~\bibnamefont {Xu}}, \ and\ \bibinfo {author}
  {\bibfnamefont {G.}~\bibnamefont {Zhu}},\ }\href@noop {} {\  (\bibinfo {year}
  {2014})},\ \Eprint {http://arxiv.org/abs/1407.4912} {arXiv:1407.4912
  [hep-ph]} \BibitemShut {NoStop}%
%%CITATION = ARXIV:1407.4912;%%
\bibitem [{\citenamefont {Ohnemus}(1991)}]{WW:NLO1}%
  \BibitemOpen
  \bibfield  {author} {\bibinfo {author} {\bibfnamefont {J.}~\bibnamefont
  {Ohnemus}},\ }\href {\doibase 10.1103/PhysRevD.44.1403} {\bibfield  {journal}
  {\bibinfo  {journal} {Phys.Rev.}\ }\textbf {\bibinfo {volume} {D44}},\
  \bibinfo {pages} {1403} (\bibinfo {year} {1991})}\BibitemShut {NoStop}%
%%CITATION = PHRVA,D44,1403;%%
\bibitem [{\citenamefont {Frixione}(1993)}]{WW:NLO2}%
  \BibitemOpen
  \bibfield  {author} {\bibinfo {author} {\bibfnamefont {S.}~\bibnamefont
  {Frixione}},\ }\href {\doibase 10.1016/0550-3213(93)90435-R} {\bibfield
  {journal} {\bibinfo  {journal} {Nucl.Phys.}\ }\textbf {\bibinfo {volume}
  {B410}},\ \bibinfo {pages} {280} (\bibinfo {year} {1993})}\BibitemShut
  {NoStop}%
%%CITATION = NUPHA,B410,280;%%
\bibitem [{\citenamefont {Ohnemus}\ and\ \citenamefont
  {Owens}(1991)}]{ZZ:NLO1}%
  \BibitemOpen
  \bibfield  {author} {\bibinfo {author} {\bibfnamefont {J.}~\bibnamefont
  {Ohnemus}}\ and\ \bibinfo {author} {\bibfnamefont {J.}~\bibnamefont
  {Owens}},\ }\href {\doibase 10.1103/PhysRevD.43.3626} {\bibfield  {journal}
  {\bibinfo  {journal} {Phys.Rev.}\ }\textbf {\bibinfo {volume} {D43}},\
  \bibinfo {pages} {3626} (\bibinfo {year} {1991})}\BibitemShut {NoStop}%
%%CITATION = PHRVA,D43,3626;%%
\bibitem [{\citenamefont {Mele}\ \emph {et~al.}(1991)\citenamefont {Mele},
  \citenamefont {Nason},\ and\ \citenamefont {Ridolfi}}]{ZZ:NLO2}%
  \BibitemOpen
  \bibfield  {author} {\bibinfo {author} {\bibfnamefont {B.}~\bibnamefont
  {Mele}}, \bibinfo {author} {\bibfnamefont {P.}~\bibnamefont {Nason}}, \ and\
  \bibinfo {author} {\bibfnamefont {G.}~\bibnamefont {Ridolfi}},\ }\href
  {\doibase 10.1016/0550-3213(91)90475-D} {\bibfield  {journal} {\bibinfo
  {journal} {Nucl.Phys.}\ }\textbf {\bibinfo {volume} {B357}},\ \bibinfo
  {pages} {409} (\bibinfo {year} {1991})}\BibitemShut {NoStop}%
%%CITATION = NUPHA,B357,409;%%
\bibitem [{\citenamefont {Frixione}\ \emph {et~al.}(1992)\citenamefont
  {Frixione}, \citenamefont {Nason},\ and\ \citenamefont {Ridolfi}}]{WZ:NLO}%
  \BibitemOpen
  \bibfield  {author} {\bibinfo {author} {\bibfnamefont {S.}~\bibnamefont
  {Frixione}}, \bibinfo {author} {\bibfnamefont {P.}~\bibnamefont {Nason}}, \
  and\ \bibinfo {author} {\bibfnamefont {G.}~\bibnamefont {Ridolfi}},\ }\href
  {\doibase 10.1016/0550-3213(92)90668-2} {\bibfield  {journal} {\bibinfo
  {journal} {Nucl.Phys.}\ }\textbf {\bibinfo {volume} {B383}},\ \bibinfo
  {pages} {3} (\bibinfo {year} {1992})}\BibitemShut {NoStop}%
%%CITATION = NUPHA,B383,3;%%
\bibitem [{\citenamefont {Ohnemus}(1994)}]{VV:Leptonic}%
  \BibitemOpen
  \bibfield  {author} {\bibinfo {author} {\bibfnamefont {J.}~\bibnamefont
  {Ohnemus}},\ }\href {\doibase 10.1103/PhysRevD.50.1931} {\bibfield  {journal}
  {\bibinfo  {journal} {Phys.Rev.}\ }\textbf {\bibinfo {volume} {D50}},\
  \bibinfo {pages} {1931} (\bibinfo {year} {1994})},\ \Eprint
  {http://arxiv.org/abs/hep-ph/9403331} {arXiv:hep-ph/9403331 [hep-ph]}
  \BibitemShut {NoStop}%
%%CITATION = HEP-PH/9403331;%%
\bibitem [{\citenamefont {Dixon}\ \emph {et~al.}(1998)\citenamefont {Dixon},
  \citenamefont {Kunszt},\ and\ \citenamefont {Signer}}]{VV:Helicity}%
  \BibitemOpen
  \bibfield  {author} {\bibinfo {author} {\bibfnamefont {L.~J.}\ \bibnamefont
  {Dixon}}, \bibinfo {author} {\bibfnamefont {Z.}~\bibnamefont {Kunszt}}, \
  and\ \bibinfo {author} {\bibfnamefont {A.}~\bibnamefont {Signer}},\ }\href
  {\doibase 10.1016/S0550-3213(98)00421-0} {\bibfield  {journal} {\bibinfo
  {journal} {Nucl.Phys.}\ }\textbf {\bibinfo {volume} {B531}},\ \bibinfo
  {pages} {3} (\bibinfo {year} {1998})},\ \Eprint
  {http://arxiv.org/abs/hep-ph/9803250} {arXiv:hep-ph/9803250 [hep-ph]}
  \BibitemShut {NoStop}%
%%CITATION = HEP-PH/9803250;%%
\bibitem [{\citenamefont {Campbell}\ and\ \citenamefont
  {Ellis}(1999)}]{VV:NLOfull1}%
  \BibitemOpen
  \bibfield  {author} {\bibinfo {author} {\bibfnamefont {J.~M.}\ \bibnamefont
  {Campbell}}\ and\ \bibinfo {author} {\bibfnamefont {R.~K.}\ \bibnamefont
  {Ellis}},\ }\href {\doibase 10.1103/PhysRevD.60.113006} {\bibfield  {journal}
  {\bibinfo  {journal} {Phys.Rev.}\ }\textbf {\bibinfo {volume} {D60}},\
  \bibinfo {pages} {113006} (\bibinfo {year} {1999})},\ \Eprint
  {http://arxiv.org/abs/hep-ph/9905386} {arXiv:hep-ph/9905386 [hep-ph]}
  \BibitemShut {NoStop}%
%%CITATION = HEP-PH/9905386;%%
\bibitem [{\citenamefont {Dixon}\ \emph {et~al.}(1999)\citenamefont {Dixon},
  \citenamefont {Kunszt},\ and\ \citenamefont {Signer}}]{VV:NLOfull2}%
  \BibitemOpen
  \bibfield  {author} {\bibinfo {author} {\bibfnamefont {L.~J.}\ \bibnamefont
  {Dixon}}, \bibinfo {author} {\bibfnamefont {Z.}~\bibnamefont {Kunszt}}, \
  and\ \bibinfo {author} {\bibfnamefont {A.}~\bibnamefont {Signer}},\ }\href
  {\doibase 10.1103/PhysRevD.60.114037} {\bibfield  {journal} {\bibinfo
  {journal} {Phys.Rev.}\ }\textbf {\bibinfo {volume} {D60}},\ \bibinfo {pages}
  {114037} (\bibinfo {year} {1999})},\ \Eprint
  {http://arxiv.org/abs/hep-ph/9907305} {arXiv:hep-ph/9907305 [hep-ph]}
  \BibitemShut {NoStop}%
%%CITATION = HEP-PH/9907305;%%
\bibitem [{\citenamefont {Glover}\ and\ \citenamefont {van~der
  Bij}(1989)}]{ggVV:1}%
  \BibitemOpen
  \bibfield  {author} {\bibinfo {author} {\bibfnamefont {E.~N.}\ \bibnamefont
  {Glover}}\ and\ \bibinfo {author} {\bibfnamefont {J.}~\bibnamefont {van~der
  Bij}},\ }\href {\doibase 10.1016/0370-2693(89)91099-X} {\bibfield  {journal}
  {\bibinfo  {journal} {Phys.Lett.}\ }\textbf {\bibinfo {volume} {B219}},\
  \bibinfo {pages} {488} (\bibinfo {year} {1989})}\BibitemShut {NoStop}%
%%CITATION = PHLTA,B219,488;%%
\bibitem [{\citenamefont {Dicus}\ \emph {et~al.}(1987)\citenamefont {Dicus},
  \citenamefont {Kao},\ and\ \citenamefont {Repko}}]{ggVV:2}%
  \BibitemOpen
  \bibfield  {author} {\bibinfo {author} {\bibfnamefont {D.~A.}\ \bibnamefont
  {Dicus}}, \bibinfo {author} {\bibfnamefont {C.}~\bibnamefont {Kao}}, \ and\
  \bibinfo {author} {\bibfnamefont {W.}~\bibnamefont {Repko}},\ }\href
  {\doibase 10.1103/PhysRevD.36.1570} {\bibfield  {journal} {\bibinfo
  {journal} {Phys.Rev.}\ }\textbf {\bibinfo {volume} {D36}},\ \bibinfo {pages}
  {1570} (\bibinfo {year} {1987})}\BibitemShut {NoStop}%
%%CITATION = PHRVA,D36,1570;%%
\bibitem [{\citenamefont {Matsuura}\ and\ \citenamefont {van~der
  Bij}(1991)}]{ggZZ:lep1}%
  \BibitemOpen
  \bibfield  {author} {\bibinfo {author} {\bibfnamefont {T.}~\bibnamefont
  {Matsuura}}\ and\ \bibinfo {author} {\bibfnamefont {J.}~\bibnamefont {van~der
  Bij}},\ }\href {\doibase 10.1007/BF01475793} {\bibfield  {journal} {\bibinfo
  {journal} {Z.Phys.}\ }\textbf {\bibinfo {volume} {C51}},\ \bibinfo {pages}
  {259} (\bibinfo {year} {1991})}\BibitemShut {NoStop}%
%%CITATION = ZEPYA,C51,259;%%
\bibitem [{\citenamefont {Zecher}\ \emph {et~al.}(1994)\citenamefont {Zecher},
  \citenamefont {Matsuura},\ and\ \citenamefont {van~der Bij}}]{ggZZ:lep2}%
  \BibitemOpen
  \bibfield  {author} {\bibinfo {author} {\bibfnamefont {C.}~\bibnamefont
  {Zecher}}, \bibinfo {author} {\bibfnamefont {T.}~\bibnamefont {Matsuura}}, \
  and\ \bibinfo {author} {\bibfnamefont {J.}~\bibnamefont {van~der Bij}},\
  }\href {\doibase 10.1007/BF01557393} {\bibfield  {journal} {\bibinfo
  {journal} {Z.Phys.}\ }\textbf {\bibinfo {volume} {C64}},\ \bibinfo {pages}
  {219} (\bibinfo {year} {1994})},\ \Eprint
  {http://arxiv.org/abs/hep-ph/9404295} {arXiv:hep-ph/9404295 [hep-ph]}
  \BibitemShut {NoStop}%
%%CITATION = HEP-PH/9404295;%%
\bibitem [{\citenamefont {Binoth}\ \emph {et~al.}(2008)\citenamefont {Binoth},
  \citenamefont {Kauer},\ and\ \citenamefont {Mertsch}}]{ggZZ:lep3}%
  \BibitemOpen
  \bibfield  {author} {\bibinfo {author} {\bibfnamefont {T.}~\bibnamefont
  {Binoth}}, \bibinfo {author} {\bibfnamefont {N.}~\bibnamefont {Kauer}}, \
  and\ \bibinfo {author} {\bibfnamefont {P.}~\bibnamefont {Mertsch}},\ }\href
  {\doibase 10.3360/dis.2008.142} {\ ,\ \bibinfo {pages} {142} (\bibinfo {year}
  {2008})},\ \Eprint {http://arxiv.org/abs/0807.0024} {arXiv:0807.0024
  [hep-ph]} \BibitemShut {NoStop}%
%%CITATION = ARXIV:0807.0024;%%
\bibitem [{\citenamefont {Binoth}\ \emph {et~al.}(2005)\citenamefont {Binoth},
  \citenamefont {Ciccolini}, \citenamefont {Kauer},\ and\ \citenamefont
  {Kramer}}]{ggWW:lep1}%
  \BibitemOpen
  \bibfield  {author} {\bibinfo {author} {\bibfnamefont {T.}~\bibnamefont
  {Binoth}}, \bibinfo {author} {\bibfnamefont {M.}~\bibnamefont {Ciccolini}},
  \bibinfo {author} {\bibfnamefont {N.}~\bibnamefont {Kauer}}, \ and\ \bibinfo
  {author} {\bibfnamefont {M.}~\bibnamefont {Kramer}},\ }\href {\doibase
  10.1088/1126-6708/2005/03/065} {\bibfield  {journal} {\bibinfo  {journal}
  {JHEP}\ }\textbf {\bibinfo {volume} {0503}},\ \bibinfo {pages} {065}
  (\bibinfo {year} {2005})},\ \Eprint {http://arxiv.org/abs/hep-ph/0503094}
  {arXiv:hep-ph/0503094 [hep-ph]} \BibitemShut {NoStop}%
%%CITATION = HEP-PH/0503094;%%
\bibitem [{\citenamefont {Binoth}\ \emph {et~al.}(2006)\citenamefont {Binoth},
  \citenamefont {Ciccolini}, \citenamefont {Kauer},\ and\ \citenamefont
  {Kramer}}]{ggWW:lep2}%
  \BibitemOpen
  \bibfield  {author} {\bibinfo {author} {\bibfnamefont {T.}~\bibnamefont
  {Binoth}}, \bibinfo {author} {\bibfnamefont {M.}~\bibnamefont {Ciccolini}},
  \bibinfo {author} {\bibfnamefont {N.}~\bibnamefont {Kauer}}, \ and\ \bibinfo
  {author} {\bibfnamefont {M.}~\bibnamefont {Kramer}},\ }\href {\doibase
  10.1088/1126-6708/2006/12/046} {\bibfield  {journal} {\bibinfo  {journal}
  {JHEP}\ }\textbf {\bibinfo {volume} {0612}},\ \bibinfo {pages} {046}
  (\bibinfo {year} {2006})},\ \Eprint {http://arxiv.org/abs/hep-ph/0611170}
  {arXiv:hep-ph/0611170 [hep-ph]} \BibitemShut {NoStop}%
%%CITATION = HEP-PH/0611170;%%
\bibitem [{\citenamefont {Campbell}\ \emph {et~al.}(2011)\citenamefont
  {Campbell}, \citenamefont {Ellis},\ and\ \citenamefont
  {Williams}}]{VV:NLOfull+gg}%
  \BibitemOpen
  \bibfield  {author} {\bibinfo {author} {\bibfnamefont {J.~M.}\ \bibnamefont
  {Campbell}}, \bibinfo {author} {\bibfnamefont {R.~K.}\ \bibnamefont {Ellis}},
  \ and\ \bibinfo {author} {\bibfnamefont {C.}~\bibnamefont {Williams}},\
  }\href {\doibase 10.1007/JHEP07(2011)018} {\bibfield  {journal} {\bibinfo
  {journal} {JHEP}\ }\textbf {\bibinfo {volume} {1107}},\ \bibinfo {pages}
  {018} (\bibinfo {year} {2011})},\ \Eprint {http://arxiv.org/abs/1105.0020}
  {arXiv:1105.0020 [hep-ph]} \BibitemShut {NoStop}%
%%CITATION = ARXIV:1105.0020;%%
\bibitem [{\citenamefont {Bierweiler}\ \emph {et~al.}(2012)\citenamefont
  {Bierweiler}, \citenamefont {Kasprzik}, \citenamefont {KŸhn},\ and\
  \citenamefont {Uccirati}}]{WW:EW1}%
  \BibitemOpen
  \bibfield  {author} {\bibinfo {author} {\bibfnamefont {A.}~\bibnamefont
  {Bierweiler}}, \bibinfo {author} {\bibfnamefont {T.}~\bibnamefont
  {Kasprzik}}, \bibinfo {author} {\bibfnamefont {J.~H.}\ \bibnamefont {KŸhn}},
  \ and\ \bibinfo {author} {\bibfnamefont {S.}~\bibnamefont {Uccirati}},\
  }\href {\doibase 10.1007/JHEP11(2012)093} {\bibfield  {journal} {\bibinfo
  {journal} {JHEP}\ }\textbf {\bibinfo {volume} {1211}},\ \bibinfo {pages}
  {093} (\bibinfo {year} {2012})},\ \Eprint {http://arxiv.org/abs/1208.3147}
  {arXiv:1208.3147 [hep-ph]} \BibitemShut {NoStop}%
%%CITATION = ARXIV:1208.3147;%%
\bibitem [{\citenamefont {Billoni}\ \emph {et~al.}(2013)\citenamefont
  {Billoni}, \citenamefont {Dittmaier}, \citenamefont {JŠger},\ and\
  \citenamefont {Speckner}}]{WW:EW2}%
  \BibitemOpen
  \bibfield  {author} {\bibinfo {author} {\bibfnamefont {M.}~\bibnamefont
  {Billoni}}, \bibinfo {author} {\bibfnamefont {S.}~\bibnamefont {Dittmaier}},
  \bibinfo {author} {\bibfnamefont {B.}~\bibnamefont {JŠger}}, \ and\ \bibinfo
  {author} {\bibfnamefont {C.}~\bibnamefont {Speckner}},\ }\href {\doibase
  10.1007/JHEP12(2013)043} {\bibfield  {journal} {\bibinfo  {journal} {JHEP}\
  }\textbf {\bibinfo {volume} {1312}},\ \bibinfo {pages} {043} (\bibinfo {year}
  {2013})},\ \Eprint {http://arxiv.org/abs/1310.1564} {arXiv:1310.1564
  [hep-ph]} \BibitemShut {NoStop}%
%%CITATION = ARXIV:1310.1564;%%
\bibitem [{\citenamefont {Bierweiler}\ \emph {et~al.}(2013)\citenamefont
  {Bierweiler}, \citenamefont {Kasprzik},\ and\ \citenamefont {KŸhn}}]{VV:EW1}%
  \BibitemOpen
  \bibfield  {author} {\bibinfo {author} {\bibfnamefont {A.}~\bibnamefont
  {Bierweiler}}, \bibinfo {author} {\bibfnamefont {T.}~\bibnamefont
  {Kasprzik}}, \ and\ \bibinfo {author} {\bibfnamefont {J.~H.}\ \bibnamefont
  {KŸhn}},\ }\href {\doibase 10.1007/JHEP12(2013)071} {\bibfield  {journal}
  {\bibinfo  {journal} {JHEP}\ }\textbf {\bibinfo {volume} {1312}},\ \bibinfo
  {pages} {071} (\bibinfo {year} {2013})},\ \Eprint
  {http://arxiv.org/abs/1305.5402} {arXiv:1305.5402 [hep-ph]} \BibitemShut
  {NoStop}%
%%CITATION = ARXIV:1305.5402;%%
\bibitem [{\citenamefont {Baglio}\ \emph {et~al.}(2013)\citenamefont {Baglio},
  \citenamefont {Ninh},\ and\ \citenamefont {Weber}}]{VV:EW2}%
  \BibitemOpen
  \bibfield  {author} {\bibinfo {author} {\bibfnamefont {J.}~\bibnamefont
  {Baglio}}, \bibinfo {author} {\bibfnamefont {L.~D.}\ \bibnamefont {Ninh}}, \
  and\ \bibinfo {author} {\bibfnamefont {M.~M.}\ \bibnamefont {Weber}},\ }\href
  {\doibase 10.1103/PhysRevD.88.113005} {\bibfield  {journal} {\bibinfo
  {journal} {Phys.Rev.}\ }\textbf {\bibinfo {volume} {D88}},\ \bibinfo {pages}
  {113005} (\bibinfo {year} {2013})},\ \Eprint {http://arxiv.org/abs/1307.4331}
  {arXiv:1307.4331} \BibitemShut {NoStop}%
%%CITATION = ARXIV:1307.4331;%%
\bibitem [{\citenamefont {Dittmaier}\ \emph {et~al.}(2008)\citenamefont
  {Dittmaier}, \citenamefont {Kallweit},\ and\ \citenamefont
  {Uwer}}]{WW+1jet:1}%
  \BibitemOpen
  \bibfield  {author} {\bibinfo {author} {\bibfnamefont {S.}~\bibnamefont
  {Dittmaier}}, \bibinfo {author} {\bibfnamefont {S.}~\bibnamefont {Kallweit}},
  \ and\ \bibinfo {author} {\bibfnamefont {P.}~\bibnamefont {Uwer}},\ }\href
  {\doibase 10.1103/PhysRevLett.100.062003} {\bibfield  {journal} {\bibinfo
  {journal} {Phys.Rev.Lett.}\ }\textbf {\bibinfo {volume} {100}},\ \bibinfo
  {pages} {062003} (\bibinfo {year} {2008})},\ \Eprint
  {http://arxiv.org/abs/0710.1577} {arXiv:0710.1577 [hep-ph]} \BibitemShut
  {NoStop}%
%%CITATION = ARXIV:0710.1577;%%
\bibitem [{\citenamefont {Campbell}\ \emph {et~al.}(2007)\citenamefont
  {Campbell}, \citenamefont {Ellis},\ and\ \citenamefont
  {Zanderighi}}]{WW+1jet:2}%
  \BibitemOpen
  \bibfield  {author} {\bibinfo {author} {\bibfnamefont {J.~M.}\ \bibnamefont
  {Campbell}}, \bibinfo {author} {\bibfnamefont {R.~K.}\ \bibnamefont {Ellis}},
  \ and\ \bibinfo {author} {\bibfnamefont {G.}~\bibnamefont {Zanderighi}},\
  }\href {\doibase 10.1088/1126-6708/2007/12/056} {\bibfield  {journal}
  {\bibinfo  {journal} {JHEP}\ }\textbf {\bibinfo {volume} {0712}},\ \bibinfo
  {pages} {056} (\bibinfo {year} {2007})},\ \Eprint
  {http://arxiv.org/abs/0710.1832} {arXiv:0710.1832 [hep-ph]} \BibitemShut
  {NoStop}%
%%CITATION = ARXIV:0710.1832;%%
\bibitem [{\citenamefont {Dittmaier}\ \emph {et~al.}(2010)\citenamefont
  {Dittmaier}, \citenamefont {Kallweit},\ and\ \citenamefont
  {Uwer}}]{WW+1jet:3}%
  \BibitemOpen
  \bibfield  {author} {\bibinfo {author} {\bibfnamefont {S.}~\bibnamefont
  {Dittmaier}}, \bibinfo {author} {\bibfnamefont {S.}~\bibnamefont {Kallweit}},
  \ and\ \bibinfo {author} {\bibfnamefont {P.}~\bibnamefont {Uwer}},\ }\href
  {\doibase 10.1016/j.nuclphysb.2009.09.029} {\bibfield  {journal} {\bibinfo
  {journal} {Nucl.Phys.}\ }\textbf {\bibinfo {volume} {B826}},\ \bibinfo
  {pages} {18} (\bibinfo {year} {2010})},\ \Eprint
  {http://arxiv.org/abs/0908.4124} {arXiv:0908.4124 [hep-ph]} \BibitemShut
  {NoStop}%
%%CITATION = ARXIV:0908.4124;%%
\bibitem [{\citenamefont {Binoth}\ \emph {et~al.}(2010)\citenamefont {Binoth},
  \citenamefont {Gleisberg}, \citenamefont {Karg}, \citenamefont {Kauer},\ and\
  \citenamefont {Sanguinetti}}]{ZZ+1jet}%
  \BibitemOpen
  \bibfield  {author} {\bibinfo {author} {\bibfnamefont {T.}~\bibnamefont
  {Binoth}}, \bibinfo {author} {\bibfnamefont {T.}~\bibnamefont {Gleisberg}},
  \bibinfo {author} {\bibfnamefont {S.}~\bibnamefont {Karg}}, \bibinfo {author}
  {\bibfnamefont {N.}~\bibnamefont {Kauer}}, \ and\ \bibinfo {author}
  {\bibfnamefont {G.}~\bibnamefont {Sanguinetti}},\ }\href {\doibase
  10.1016/j.physletb.2009.12.013} {\bibfield  {journal} {\bibinfo  {journal}
  {Phys.Lett.}\ }\textbf {\bibinfo {volume} {B683}},\ \bibinfo {pages} {154}
  (\bibinfo {year} {2010})},\ \Eprint {http://arxiv.org/abs/0911.3181}
  {arXiv:0911.3181 [hep-ph]} \BibitemShut {NoStop}%
%%CITATION = ARXIV:0911.3181;%%
\bibitem [{\citenamefont {Melia}\ \emph {et~al.}(2011)\citenamefont {Melia},
  \citenamefont {Melnikov}, \citenamefont {Rontsch},\ and\ \citenamefont
  {Zanderighi}}]{WW+2jets:1}%
  \BibitemOpen
  \bibfield  {author} {\bibinfo {author} {\bibfnamefont {T.}~\bibnamefont
  {Melia}}, \bibinfo {author} {\bibfnamefont {K.}~\bibnamefont {Melnikov}},
  \bibinfo {author} {\bibfnamefont {R.}~\bibnamefont {Rontsch}}, \ and\
  \bibinfo {author} {\bibfnamefont {G.}~\bibnamefont {Zanderighi}},\ }\href
  {\doibase 10.1103/PhysRevD.83.114043} {\bibfield  {journal} {\bibinfo
  {journal} {Phys.Rev.}\ }\textbf {\bibinfo {volume} {D83}},\ \bibinfo {pages}
  {114043} (\bibinfo {year} {2011})},\ \Eprint {http://arxiv.org/abs/1104.2327}
  {arXiv:1104.2327 [hep-ph]} \BibitemShut {NoStop}%
%%CITATION = ARXIV:1104.2327;%%
\bibitem [{\citenamefont {Greiner}\ \emph {et~al.}(2012)\citenamefont
  {Greiner}, \citenamefont {Heinrich}, \citenamefont {Mastrolia}, \citenamefont
  {Ossola}, \citenamefont {Reiter} \emph {et~al.}}]{WW+2jets:2}%
  \BibitemOpen
  \bibfield  {author} {\bibinfo {author} {\bibfnamefont {N.}~\bibnamefont
  {Greiner}}, \bibinfo {author} {\bibfnamefont {G.}~\bibnamefont {Heinrich}},
  \bibinfo {author} {\bibfnamefont {P.}~\bibnamefont {Mastrolia}}, \bibinfo
  {author} {\bibfnamefont {G.}~\bibnamefont {Ossola}}, \bibinfo {author}
  {\bibfnamefont {T.}~\bibnamefont {Reiter}},  \emph {et~al.},\ }\href
  {\doibase 10.1016/j.physletb.2012.06.027} {\bibfield  {journal} {\bibinfo
  {journal} {Phys.Lett.}\ }\textbf {\bibinfo {volume} {B713}},\ \bibinfo
  {pages} {277} (\bibinfo {year} {2012})},\ \Eprint
  {http://arxiv.org/abs/1202.6004} {arXiv:1202.6004 [hep-ph]} \BibitemShut
  {NoStop}%
%%CITATION = ARXIV:1202.6004;%%
\bibitem [{\citenamefont {Wang}\ \emph {et~al.}(2013)\citenamefont {Wang},
  \citenamefont {Li}, \citenamefont {Liu}, \citenamefont {Shao},\ and\
  \citenamefont {Li}}]{VV:pT}%
  \BibitemOpen
  \bibfield  {author} {\bibinfo {author} {\bibfnamefont {Y.}~\bibnamefont
  {Wang}}, \bibinfo {author} {\bibfnamefont {C.~S.}\ \bibnamefont {Li}},
  \bibinfo {author} {\bibfnamefont {Z.~L.}\ \bibnamefont {Liu}}, \bibinfo
  {author} {\bibfnamefont {D.~Y.}\ \bibnamefont {Shao}}, \ and\ \bibinfo
  {author} {\bibfnamefont {H.~T.}\ \bibnamefont {Li}},\ }\href {\doibase
  10.1103/PhysRevD.88.114017} {\bibfield  {journal} {\bibinfo  {journal}
  {Phys.Rev.}\ }\textbf {\bibinfo {volume} {D88}},\ \bibinfo {pages} {114017}
  (\bibinfo {year} {2013})},\ \Eprint {http://arxiv.org/abs/1307.7520}
  {arXiv:1307.7520} \BibitemShut {NoStop}%
%%CITATION = ARXIV:1307.7520;%%
\bibitem [{\citenamefont {Grazzini}(2006)}]{WW:pT1}%
  \BibitemOpen
  \bibfield  {author} {\bibinfo {author} {\bibfnamefont {M.}~\bibnamefont
  {Grazzini}},\ }\href {\doibase 10.1088/1126-6708/2006/01/095} {\bibfield
  {journal} {\bibinfo  {journal} {JHEP}\ }\textbf {\bibinfo {volume} {0601}},\
  \bibinfo {pages} {095} (\bibinfo {year} {2006})},\ \Eprint
  {http://arxiv.org/abs/hep-ph/0510337} {arXiv:hep-ph/0510337 [hep-ph]}
  \BibitemShut {NoStop}%
%%CITATION = HEP-PH/0510337;%%
\bibitem [{\citenamefont {Meade}\ \emph {et~al.}(2014)\citenamefont {Meade},
  \citenamefont {Ramani},\ and\ \citenamefont {Zeng}}]{WW:pT2}%
  \BibitemOpen
  \bibfield  {author} {\bibinfo {author} {\bibfnamefont {P.}~\bibnamefont
  {Meade}}, \bibinfo {author} {\bibfnamefont {H.}~\bibnamefont {Ramani}}, \
  and\ \bibinfo {author} {\bibfnamefont {M.}~\bibnamefont {Zeng}},\ }\href@noop
  {} {\  (\bibinfo {year} {2014})},\ \Eprint {http://arxiv.org/abs/1407.4481}
  {arXiv:1407.4481 [hep-ph]} \BibitemShut {NoStop}%
%%CITATION = ARXIV:1407.4481;%%
\bibitem [{\citenamefont {Jaiswal}\ and\ \citenamefont
  {Okui}(2014)}]{WW:JetVeto}%
  \BibitemOpen
  \bibfield  {author} {\bibinfo {author} {\bibfnamefont {P.}~\bibnamefont
  {Jaiswal}}\ and\ \bibinfo {author} {\bibfnamefont {T.}~\bibnamefont {Okui}},\
  }\href@noop {} {\  (\bibinfo {year} {2014})},\ \Eprint
  {http://arxiv.org/abs/1407.4537} {arXiv:1407.4537 [hep-ph]} \BibitemShut
  {NoStop}%
%%CITATION = ARXIV:1407.4537;%%
\bibitem [{\citenamefont {Dawson}\ \emph {et~al.}(2013)\citenamefont {Dawson},
  \citenamefont {Lewis},\ and\ \citenamefont {Zeng}}]{WW:Threshold}%
  \BibitemOpen
  \bibfield  {author} {\bibinfo {author} {\bibfnamefont {S.}~\bibnamefont
  {Dawson}}, \bibinfo {author} {\bibfnamefont {I.~M.}\ \bibnamefont {Lewis}}, \
  and\ \bibinfo {author} {\bibfnamefont {M.}~\bibnamefont {Zeng}},\ }\href
  {\doibase 10.1103/PhysRevD.88.054028} {\bibfield  {journal} {\bibinfo
  {journal} {Phys.Rev.}\ }\textbf {\bibinfo {volume} {D88}},\ \bibinfo {pages}
  {054028} (\bibinfo {year} {2013})},\ \Eprint {http://arxiv.org/abs/1307.3249}
  {arXiv:1307.3249} \BibitemShut {NoStop}%
%%CITATION = ARXIV:1307.3249;%%
\bibitem [{\citenamefont {Wang}\ \emph {et~al.}(2014)\citenamefont {Wang},
  \citenamefont {Li}, \citenamefont {Liu},\ and\ \citenamefont
  {Shao}}]{VZ:Threshold}%
  \BibitemOpen
  \bibfield  {author} {\bibinfo {author} {\bibfnamefont {Y.}~\bibnamefont
  {Wang}}, \bibinfo {author} {\bibfnamefont {C.~S.}\ \bibnamefont {Li}},
  \bibinfo {author} {\bibfnamefont {Z.~L.}\ \bibnamefont {Liu}}, \ and\
  \bibinfo {author} {\bibfnamefont {D.~Y.}\ \bibnamefont {Shao}},\ }\href
  {\doibase 10.1103/PhysRevD.90.034008} {\bibfield  {journal} {\bibinfo
  {journal} {Phys.Rev.}\ }\textbf {\bibinfo {volume} {D90}},\ \bibinfo {pages}
  {034008} (\bibinfo {year} {2014})},\ \Eprint {http://arxiv.org/abs/1406.1417}
  {arXiv:1406.1417 [hep-ph]} \BibitemShut {NoStop}%
%%CITATION = ARXIV:1406.1417;%%
\bibitem [{\citenamefont {Gehrmann}\ \emph {et~al.}(2014)\citenamefont
  {Gehrmann}, \citenamefont {Grazzini}, \citenamefont {Kallweit}, \citenamefont
  {Maierhšfer}, \citenamefont {von Manteuffel} \emph {et~al.}}]{WW:NNLO}%
  \BibitemOpen
  \bibfield  {author} {\bibinfo {author} {\bibfnamefont {T.}~\bibnamefont
  {Gehrmann}}, \bibinfo {author} {\bibfnamefont {M.}~\bibnamefont {Grazzini}},
  \bibinfo {author} {\bibfnamefont {S.}~\bibnamefont {Kallweit}}, \bibinfo
  {author} {\bibfnamefont {P.}~\bibnamefont {Maierhšfer}}, \bibinfo {author}
  {\bibfnamefont {A.}~\bibnamefont {von Manteuffel}},  \emph {et~al.},\
  }\href@noop {} {\  (\bibinfo {year} {2014})},\ \Eprint
  {http://arxiv.org/abs/1408.5243} {arXiv:1408.5243 [hep-ph]} \BibitemShut
  {NoStop}%
%%CITATION = ARXIV:1408.5243;%%
\bibitem [{\citenamefont {Cascioli}\ \emph {et~al.}(2014)\citenamefont
  {Cascioli}, \citenamefont {Gehrmann}, \citenamefont {Grazzini}, \citenamefont
  {Kallweit}, \citenamefont {Maierhšfer} \emph {et~al.}}]{ZZ:NNLO}%
  \BibitemOpen
  \bibfield  {author} {\bibinfo {author} {\bibfnamefont {F.}~\bibnamefont
  {Cascioli}}, \bibinfo {author} {\bibfnamefont {T.}~\bibnamefont {Gehrmann}},
  \bibinfo {author} {\bibfnamefont {M.}~\bibnamefont {Grazzini}}, \bibinfo
  {author} {\bibfnamefont {S.}~\bibnamefont {Kallweit}}, \bibinfo {author}
  {\bibfnamefont {P.}~\bibnamefont {Maierhšfer}},  \emph {et~al.},\ }\href
  {\doibase 10.1016/j.physletb.2014.06.056} {\bibfield  {journal} {\bibinfo
  {journal} {Phys.Lett.}\ }\textbf {\bibinfo {volume} {B735}},\ \bibinfo
  {pages} {311} (\bibinfo {year} {2014})},\ \Eprint
  {http://arxiv.org/abs/1405.2219} {arXiv:1405.2219 [hep-ph]} \BibitemShut
  {NoStop}%
%%CITATION = ARXIV:1405.2219;%%
\bibitem [{\citenamefont {Caola}\ \emph {et~al.}(2014)\citenamefont {Caola},
  \citenamefont {Henn}, \citenamefont {Melnikov}, \citenamefont {Smirnov},\
  and\ \citenamefont {Smirnov}}]{VV:NNLO}%
  \BibitemOpen
  \bibfield  {author} {\bibinfo {author} {\bibfnamefont {F.}~\bibnamefont
  {Caola}}, \bibinfo {author} {\bibfnamefont {J.~M.}\ \bibnamefont {Henn}},
  \bibinfo {author} {\bibfnamefont {K.}~\bibnamefont {Melnikov}}, \bibinfo
  {author} {\bibfnamefont {A.~V.}\ \bibnamefont {Smirnov}}, \ and\ \bibinfo
  {author} {\bibfnamefont {V.~A.}\ \bibnamefont {Smirnov}},\ }\href@noop {} {\
  (\bibinfo {year} {2014})},\ \Eprint {http://arxiv.org/abs/1408.6409}
  {arXiv:1408.6409 [hep-ph]} \BibitemShut {NoStop}%
%%CITATION = ARXIV:1408.6409;%%
\bibitem [{\citenamefont {Parisi}(1980)}]{PiSq:1}%
  \BibitemOpen
  \bibfield  {author} {\bibinfo {author} {\bibfnamefont {G.}~\bibnamefont
  {Parisi}},\ }\href {\doibase 10.1016/0370-2693(80)90746-7} {\bibfield
  {journal} {\bibinfo  {journal} {Phys.Lett.}\ }\textbf {\bibinfo {volume}
  {B90}},\ \bibinfo {pages} {295} (\bibinfo {year} {1980})}\BibitemShut
  {NoStop}%
%%CITATION = PHLTA,B90,295;%%
\bibitem [{\citenamefont {Sterman}(1987)}]{PiSq:2}%
  \BibitemOpen
  \bibfield  {author} {\bibinfo {author} {\bibfnamefont {G.~F.}\ \bibnamefont
  {Sterman}},\ }\href {\doibase 10.1016/0550-3213(87)90258-6} {\bibfield
  {journal} {\bibinfo  {journal} {Nucl.Phys.}\ }\textbf {\bibinfo {volume}
  {B281}},\ \bibinfo {pages} {310} (\bibinfo {year} {1987})}\BibitemShut
  {NoStop}%
%%CITATION = NUPHA,B281,310;%%
\bibitem [{\citenamefont {Magnea}\ and\ \citenamefont
  {Sterman}(1990)}]{PiSq:3}%
  \BibitemOpen
  \bibfield  {author} {\bibinfo {author} {\bibfnamefont {L.}~\bibnamefont
  {Magnea}}\ and\ \bibinfo {author} {\bibfnamefont {G.~F.}\ \bibnamefont
  {Sterman}},\ }\href {\doibase 10.1103/PhysRevD.42.4222} {\bibfield  {journal}
  {\bibinfo  {journal} {Phys.Rev.}\ }\textbf {\bibinfo {volume} {D42}},\
  \bibinfo {pages} {4222} (\bibinfo {year} {1990})}\BibitemShut {NoStop}%
%%CITATION = PHRVA,D42,4222;%%
\bibitem [{\citenamefont {Eynck}\ \emph {et~al.}(2003)\citenamefont {Eynck},
  \citenamefont {Laenen},\ and\ \citenamefont {Magnea}}]{PiSq:4}%
  \BibitemOpen
  \bibfield  {author} {\bibinfo {author} {\bibfnamefont {T.~O.}\ \bibnamefont
  {Eynck}}, \bibinfo {author} {\bibfnamefont {E.}~\bibnamefont {Laenen}}, \
  and\ \bibinfo {author} {\bibfnamefont {L.}~\bibnamefont {Magnea}},\ }\href
  {\doibase 10.1088/1126-6708/2003/06/057} {\bibfield  {journal} {\bibinfo
  {journal} {JHEP}\ }\textbf {\bibinfo {volume} {0306}},\ \bibinfo {pages}
  {057} (\bibinfo {year} {2003})},\ \Eprint
  {http://arxiv.org/abs/hep-ph/0305179} {arXiv:hep-ph/0305179 [hep-ph]}
  \BibitemShut {NoStop}%
%%CITATION = HEP-PH/0305179;%%
\bibitem [{\citenamefont {Ahrens}\ \emph {et~al.}(2009)\citenamefont {Ahrens},
  \citenamefont {Becher}, \citenamefont {Neubert},\ and\ \citenamefont
  {Yang}}]{PiSq:Higgs}%
  \BibitemOpen
  \bibfield  {author} {\bibinfo {author} {\bibfnamefont {V.}~\bibnamefont
  {Ahrens}}, \bibinfo {author} {\bibfnamefont {T.}~\bibnamefont {Becher}},
  \bibinfo {author} {\bibfnamefont {M.}~\bibnamefont {Neubert}}, \ and\
  \bibinfo {author} {\bibfnamefont {L.~L.}\ \bibnamefont {Yang}},\ }\href
  {\doibase 10.1103/PhysRevD.79.033013} {\bibfield  {journal} {\bibinfo
  {journal} {Phys.Rev.}\ }\textbf {\bibinfo {volume} {D79}},\ \bibinfo {pages}
  {033013} (\bibinfo {year} {2009})},\ \Eprint {http://arxiv.org/abs/0808.3008}
  {arXiv:0808.3008 [hep-ph]} \BibitemShut {NoStop}%
%%CITATION = ARXIV:0808.3008;%%
\bibitem [{\citenamefont {Bauer}\ \emph {et~al.}(2000)\citenamefont {Bauer},
  \citenamefont {Fleming},\ and\ \citenamefont {Luke}}]{SCET:label1}%
  \BibitemOpen
  \bibfield  {author} {\bibinfo {author} {\bibfnamefont {C.~W.}\ \bibnamefont
  {Bauer}}, \bibinfo {author} {\bibfnamefont {S.}~\bibnamefont {Fleming}}, \
  and\ \bibinfo {author} {\bibfnamefont {M.~E.}\ \bibnamefont {Luke}},\ }\href
  {\doibase 10.1103/PhysRevD.63.014006} {\bibfield  {journal} {\bibinfo
  {journal} {Phys.Rev.}\ }\textbf {\bibinfo {volume} {D63}},\ \bibinfo {pages}
  {014006} (\bibinfo {year} {2000})},\ \Eprint
  {http://arxiv.org/abs/hep-ph/0005275} {arXiv:hep-ph/0005275 [hep-ph]}
  \BibitemShut {NoStop}%
%%CITATION = HEP-PH/0005275;%%
\bibitem [{\citenamefont {Bauer}\ \emph {et~al.}(2001)\citenamefont {Bauer},
  \citenamefont {Fleming}, \citenamefont {Pirjol},\ and\ \citenamefont
  {Stewart}}]{SCET:label2}%
  \BibitemOpen
  \bibfield  {author} {\bibinfo {author} {\bibfnamefont {C.~W.}\ \bibnamefont
  {Bauer}}, \bibinfo {author} {\bibfnamefont {S.}~\bibnamefont {Fleming}},
  \bibinfo {author} {\bibfnamefont {D.}~\bibnamefont {Pirjol}}, \ and\ \bibinfo
  {author} {\bibfnamefont {I.~W.}\ \bibnamefont {Stewart}},\ }\href {\doibase
  10.1103/PhysRevD.63.114020} {\bibfield  {journal} {\bibinfo  {journal}
  {Phys.Rev.}\ }\textbf {\bibinfo {volume} {D63}},\ \bibinfo {pages} {114020}
  (\bibinfo {year} {2001})},\ \Eprint {http://arxiv.org/abs/hep-ph/0011336}
  {arXiv:hep-ph/0011336 [hep-ph]} \BibitemShut {NoStop}%
%%CITATION = HEP-PH/0011336;%%
\bibitem [{\citenamefont {Bauer}\ and\ \citenamefont
  {Stewart}(2001)}]{SCET:label3}%
  \BibitemOpen
  \bibfield  {author} {\bibinfo {author} {\bibfnamefont {C.~W.}\ \bibnamefont
  {Bauer}}\ and\ \bibinfo {author} {\bibfnamefont {I.~W.}\ \bibnamefont
  {Stewart}},\ }\href {\doibase 10.1016/S0370-2693(01)00902-9} {\bibfield
  {journal} {\bibinfo  {journal} {Phys.Lett.}\ }\textbf {\bibinfo {volume}
  {B516}},\ \bibinfo {pages} {134} (\bibinfo {year} {2001})},\ \Eprint
  {http://arxiv.org/abs/hep-ph/0107001} {arXiv:hep-ph/0107001 [hep-ph]}
  \BibitemShut {NoStop}%
%%CITATION = HEP-PH/0107001;%%
\bibitem [{\citenamefont {Bauer}\ \emph {et~al.}(2002)\citenamefont {Bauer},
  \citenamefont {Pirjol},\ and\ \citenamefont {Stewart}}]{SCET:label4}%
  \BibitemOpen
  \bibfield  {author} {\bibinfo {author} {\bibfnamefont {C.~W.}\ \bibnamefont
  {Bauer}}, \bibinfo {author} {\bibfnamefont {D.}~\bibnamefont {Pirjol}}, \
  and\ \bibinfo {author} {\bibfnamefont {I.~W.}\ \bibnamefont {Stewart}},\
  }\href {\doibase 10.1103/PhysRevD.65.054022} {\bibfield  {journal} {\bibinfo
  {journal} {Phys.Rev.}\ }\textbf {\bibinfo {volume} {D65}},\ \bibinfo {pages}
  {054022} (\bibinfo {year} {2002})},\ \Eprint
  {http://arxiv.org/abs/hep-ph/0109045} {arXiv:hep-ph/0109045 [hep-ph]}
  \BibitemShut {NoStop}%
%%CITATION = HEP-PH/0109045;%%
\bibitem [{\citenamefont {Beneke}\ \emph {et~al.}(2002)\citenamefont {Beneke},
  \citenamefont {Chapovsky}, \citenamefont {Diehl},\ and\ \citenamefont
  {Feldmann}}]{SCET:mp1}%
  \BibitemOpen
  \bibfield  {author} {\bibinfo {author} {\bibfnamefont {M.}~\bibnamefont
  {Beneke}}, \bibinfo {author} {\bibfnamefont {A.}~\bibnamefont {Chapovsky}},
  \bibinfo {author} {\bibfnamefont {M.}~\bibnamefont {Diehl}}, \ and\ \bibinfo
  {author} {\bibfnamefont {T.}~\bibnamefont {Feldmann}},\ }\href {\doibase
  10.1016/S0550-3213(02)00687-9} {\bibfield  {journal} {\bibinfo  {journal}
  {Nucl.Phys.}\ }\textbf {\bibinfo {volume} {B643}},\ \bibinfo {pages} {431}
  (\bibinfo {year} {2002})},\ \Eprint {http://arxiv.org/abs/hep-ph/0206152}
  {arXiv:hep-ph/0206152 [hep-ph]} \BibitemShut {NoStop}%
%%CITATION = HEP-PH/0206152;%%
\bibitem [{\citenamefont {Beneke}\ and\ \citenamefont
  {Feldmann}(2003)}]{SCET:mp2}%
  \BibitemOpen
  \bibfield  {author} {\bibinfo {author} {\bibfnamefont {M.}~\bibnamefont
  {Beneke}}\ and\ \bibinfo {author} {\bibfnamefont {T.}~\bibnamefont
  {Feldmann}},\ }\href {\doibase 10.1016/S0370-2693(02)03204-5} {\bibfield
  {journal} {\bibinfo  {journal} {Phys.Lett.}\ }\textbf {\bibinfo {volume}
  {B553}},\ \bibinfo {pages} {267} (\bibinfo {year} {2003})},\ \Eprint
  {http://arxiv.org/abs/hep-ph/0211358} {arXiv:hep-ph/0211358 [hep-ph]}
  \BibitemShut {NoStop}%
%%CITATION = HEP-PH/0211358;%%
\bibitem [{\citenamefont {Stewart}\ \emph
  {et~al.}(2010{\natexlab{a}})\citenamefont {Stewart}, \citenamefont
  {Tackmann},\ and\ \citenamefont {Waalewijn}}]{Beam:1}%
  \BibitemOpen
  \bibfield  {author} {\bibinfo {author} {\bibfnamefont {I.~W.}\ \bibnamefont
  {Stewart}}, \bibinfo {author} {\bibfnamefont {F.~J.}\ \bibnamefont
  {Tackmann}}, \ and\ \bibinfo {author} {\bibfnamefont {W.~J.}\ \bibnamefont
  {Waalewijn}},\ }\href {\doibase 10.1103/PhysRevD.81.094035} {\bibfield
  {journal} {\bibinfo  {journal} {Phys.Rev.}\ }\textbf {\bibinfo {volume}
  {D81}},\ \bibinfo {pages} {094035} (\bibinfo {year} {2010}{\natexlab{a}})},\
  \Eprint {http://arxiv.org/abs/0910.0467} {arXiv:0910.0467 [hep-ph]}
  \BibitemShut {NoStop}%
%%CITATION = ARXIV:0910.0467;%%
\bibitem [{\citenamefont {Stewart}\ \emph
  {et~al.}(2010{\natexlab{b}})\citenamefont {Stewart}, \citenamefont
  {Tackmann},\ and\ \citenamefont {Waalewijn}}]{Beam:2}%
  \BibitemOpen
  \bibfield  {author} {\bibinfo {author} {\bibfnamefont {I.~W.}\ \bibnamefont
  {Stewart}}, \bibinfo {author} {\bibfnamefont {F.~J.}\ \bibnamefont
  {Tackmann}}, \ and\ \bibinfo {author} {\bibfnamefont {W.~J.}\ \bibnamefont
  {Waalewijn}},\ }\href {\doibase 10.1007/JHEP09(2010)005} {\bibfield
  {journal} {\bibinfo  {journal} {JHEP}\ }\textbf {\bibinfo {volume} {1009}},\
  \bibinfo {pages} {005} (\bibinfo {year} {2010}{\natexlab{b}})},\ \Eprint
  {http://arxiv.org/abs/1002.2213} {arXiv:1002.2213 [hep-ph]} \BibitemShut
  {NoStop}%
%%CITATION = ARXIV:1002.2213;%%
\bibitem [{\citenamefont {Becher}\ and\ \citenamefont
  {Neubert}(2011)}]{SCET:DY}%
  \BibitemOpen
  \bibfield  {author} {\bibinfo {author} {\bibfnamefont {T.}~\bibnamefont
  {Becher}}\ and\ \bibinfo {author} {\bibfnamefont {M.}~\bibnamefont
  {Neubert}},\ }\href {\doibase 10.1140/epjc/s10052-011-1665-7} {\bibfield
  {journal} {\bibinfo  {journal} {Eur.Phys.J.}\ }\textbf {\bibinfo {volume}
  {C71}},\ \bibinfo {pages} {1665} (\bibinfo {year} {2011})},\ \Eprint
  {http://arxiv.org/abs/1007.4005} {arXiv:1007.4005 [hep-ph]} \BibitemShut
  {NoStop}%
%%CITATION = ARXIV:1007.4005;%%
\bibitem [{\citenamefont {Martin}\ \emph {et~al.}(2009)\citenamefont {Martin},
  \citenamefont {Stirling}, \citenamefont {Thorne},\ and\ \citenamefont
  {Watt}}]{PDF:MSTW2008}%
  \BibitemOpen
  \bibfield  {author} {\bibinfo {author} {\bibfnamefont {A.}~\bibnamefont
  {Martin}}, \bibinfo {author} {\bibfnamefont {W.}~\bibnamefont {Stirling}},
  \bibinfo {author} {\bibfnamefont {R.}~\bibnamefont {Thorne}}, \ and\ \bibinfo
  {author} {\bibfnamefont {G.}~\bibnamefont {Watt}},\ }\href {\doibase
  10.1140/epjc/s10052-009-1072-5} {\bibfield  {journal} {\bibinfo  {journal}
  {Eur.Phys.J.}\ }\textbf {\bibinfo {volume} {C63}},\ \bibinfo {pages} {189}
  (\bibinfo {year} {2009})},\ \Eprint {http://arxiv.org/abs/0901.0002}
  {arXiv:0901.0002 [hep-ph]} \BibitemShut {NoStop}%
%%CITATION = ARXIV:0901.0002;%%
\bibitem [{\citenamefont {Rainbolt}\ \emph {et~al.}(2014)\citenamefont
  {Rainbolt}, \citenamefont {Gunter},\ and\ \citenamefont
  {Schmitt}}]{WW:JetVeto2}%
  \BibitemOpen
  \bibfield  {author} {\bibinfo {author} {\bibfnamefont {J.~L.}\ \bibnamefont
  {Rainbolt}}, \bibinfo {author} {\bibfnamefont {T.}~\bibnamefont {Gunter}}, \
  and\ \bibinfo {author} {\bibfnamefont {M.}~\bibnamefont {Schmitt}},\
  }\href@noop {} {\  (\bibinfo {year} {2014})},\ \Eprint
  {http://arxiv.org/abs/1410.8058} {arXiv:1410.8058 [hep-ex]} \BibitemShut
  {NoStop}%
%%CITATION = ARXIV:1410.8058;%%
\end{thebibliography}%

\end{document}